%% file: ajreserving.tex
\documentclass[11pt]{article}

\input{preamble}   

\begin{document}
\title{Individual claims reserving using the Aalen--Johansen estimator}
 \author{Martin Bladt}
 \affil{Department for Mathematical Sciences, University of Copenhagen}
 \author{Gabriele Pittarello}
 \affil{Department of Statistical Sciences, Sapienza, University of Rome}
\date{\today}
 \maketitle
 \begin{abstract}

We propose an individual claims reserving model based on the conditional Aalen--Johansen estimator, as developed in \citet{bladt23}. In our approach, we formulate a multi-state problem, where the underlying variable is the individual claim size, rather than time. The states in this model represent development periods, and we estimate the cumulative density function of individual claim sizes using the conditional Aalen--Johansen method as transition probabilities to an absorbing state. Our methodology reinterprets the concept of multi-state models and offers a strategy for modeling the complete curve of individual claim sizes. To illustrate our approach, we apply our model to both simulated and real datasets. Having access to the entire dataset enables us to support the use of our approach by comparing the predicted total final cost with the actual amount, as well as evaluating it in terms of the continuously ranked probability score.
\end{abstract}
\noindent \textit{Keywords}: multi-state models; conditional Aalen--Johansen; reserving; non-life insurance.
\section{Introduction}

This manuscript demonstrates that multi-state models provide a natural framework for modeling the cost of individual claims in an individual reserving application. Our approach builds upon the findings presented in a recent paper \citep{bladt23}, which introduces the conditional Aalen--Johansen estimator. This estimator is a versatile, non-parametric tool for estimating state occupation probabilities conditioned on specific features, and it also discusses its key properties. Note that Aalen--Johansen-type estimators have traditionally found applications in life insurance. However, in this work, we demonstrate how these estimators can also be applied to non-life insurance for reserving purposes.

As mentioned earlier, multi-state models have found widespread use in the domain of life insurance. In the works of \citet{hoem69, hoem72}, they introduce a by-now standard approach, representing biometric states (active, deceased, etc.) as finite states within a spatial model. However, to the best of our knowledge, these models have seen less frequent application in non-life insurance literature.
Notable exceptions can be found in the work of \citet{hesselager94} and \citet{maciak22}. In \citet{hesselager94}, the authors model outstanding claim amounts conditionally on the current state, such as Incurred But Not Reported (IBNR), Reported But Not Settled (RBNS), and Settled. In this context, the time variable corresponds to calendar time, similar to that in life insurance. In contrast, our approach in this paper takes a different perspective, replacing time with payments as the fundamental evolution variable. In \citet{maciak22}, the authors propose modeling the incremental paid amounts as outcomes of a homogeneous Markov chain on a finite state space, though only for aggregated data and in discrete time.

In this manuscript, we propose an individual claims reserving model based on a continuous-time non-explosive pure jump process denoted by $J$ on a finite state space, $\mathcal{S} = \{1, \ldots, k\}$, $k\in\amsmathbb{N}$, where states correspond to the development periods (DP's) within a development triangle, and the ``time spent'' between state transitions corresponds to the claim size growth between DP's. The state $k$ serves as an absorbing state, representing claim closure. It is worth noting that, in practice, DP's have a stochastic behavior that leads to variation in the state space dimension across individual claim histories. However, for simplicity, we address this by collapsing long claim developments into a single state, as the specific details are not of primary importance in this context.

In a broader context, the application of counting processes to individual reserving has been a subject of research since the seminal contributions \citet{arjas89}; \citet{norberg93, norberg97}, with subsequent refinements and practical considerations detailed in the work of \citet{haastrup96} and \citet{antonio14}. In the study \citet{antonio14}, they assume a framework where claims are generated by a position-dependent marked Poisson process, and they incorporate a severity modeling approach within the mark.

The primary contribution of this manuscript lies in presenting a novel interpretation of state spaces, which enables us to approach the claims reserving problem from a fresh perspective, using a non-parametric estimator from survival analysis. We denote state process by $J_z$ when the cumulative claim size of an individual reaches $z$, where $z \in \left[0,+\infty\right)$. Note that in this approach, the fundamental variable of interest is the claim size itself, rather than time. The transition probabilities leading to the absorbing state $k$ represent the cumulative distribution of individual claim sizes. In other words, as individuals progress through the states, their observations accumulate claim size, reflecting the evolution of their claims. Upon reaching the absorbing state, a claim is considered settled, and no further payments are possible.

Recent literature on individual reserving has focused on decomposing individual reserving data into different modules that describe the micro-level structure of the portfolio, e.g. payment delay, settlement delay, payment size. By doing so, \citet{huang15} and \citet{wang21} extended the discrete model for predicting IBNR claims in \citet{norberg86} to a model for RBNS and IBNR claims that discusses, theoretically and with empirical studies, the beneficial effect of including individual information in reserving models. An interesting approach to modeling the joint distribution of reporting delays and claim sizes of RBNS claims under censoring can be found in \citet{lopez19b}. The work in \citet{crevecoeur23} embeds \citet{robben22} in a more general context and illustrates how pricing and reserving can be modelled under a unified framework and severity is modelled conditionally on the previous modules. A similar approach has been proposed in \citet{delong22}, wherein the authors incorporate a gamma-distributed severity component into a neural network-based methodology.

In contrast to the aforementioned works, which rely on likelihood-based estimation or parametric assumptions on the claim size distribution, we propose a fully non-parametric estimation of the full cumulative distribution function of claim sizes, building on the results of \citet{bladt23}. The latter reference proposes a conditional version of the classic Aalen--Johansen estimator, which allows for continuous or discrete covariates, and where minimal assumptions on the process $J$ are imposed, i.e. no Markov assumption is required. In particular, such lax probabilistic framework is desirable when it is unclear whether there exist duration effects in each state, as is the case for claim development processes.

Finally, we also propose a solution for modeling the IBNR and RBNS reserves separately. Our model is compared with the chain ladder in two case studies, one using simulated data and the other using a recent dataset from a Danish non-life insurer. In both the simulation case study and the real data case application, we have access to the total future claims costs and can thus compare the predicted amount with the target amount. The predicted curve for the cumulative density function of claims size is compared with the empirical cumulative density function using the Continuously Ranked Probability Score (CRPS), first proposed in \citet{gneiting07}. Interestingly, we show how the CRPS can be used to perform model selection by answering the natural question of how to select the number of states in the state space $\mathcal{S}$. 

This manuscript is organised as follows.
In \Cref{sec:msm}, we introduce the conditional Aalen-Johansen estimator after a discussion on the probabilistic setup and we explain how to reinterpret a state space $\mathcal{S}$ to set up our reserving model. In \Cref{ss:ibnrrbns}, we connect our framework to development triangles and after a discussion on the current literature in reserving, we illustrate how to estimate of the final cost of the claims. \Cref{s:simulation} will demonstrate the effectiveness of our approach through a comprehensive case study using simulated data. Recent contributions to the literature have provided various algorithms for generating individual loss reserve datasets \citep{avanzi21, avanzi23, gabrielli18, wang22}. Once reporting delays become available, it is possible to simulate both Incurred But Not Reported (IBNR) claims and Reported But Not Settled (RBNS) claims. We have chosen to focus on the exploration of RBNS claims in the next section, deliberately avoiding additional assumptions about the claims generation process (see for example \citet{avanzi21}). This approach aims to draw the attention of the reader to the applicability of our models in different scenarios. We showcase the application of our IBNR modeling strategy in the section \Cref{s:realdata}, where we rigorously test our models using a recent real individual data portfolio from a Danish insurer. Finally, Section \ref{sec:conc} concludes.


\section{The conditional Aalen--Johansen {\color{black} estimator}}
\label{sec:msm}

In this section we introduce our statistical framework, and the quantities of interest, the conditional occupation probabilities. We present the conditional Aalen-Johansen {\color{black} estimator \citep{bladt23}}, a non parametric estimator for the conditional occupation probabilities of an arbitrary jump process. Note that the conditional Aalen-Johansen {\color{black} estimator} is closely linked with the literature streams on kernel-based estimation for survival and semi-Markov processes {\color{black} \citep{mckeague90, nielsen95, dabrowska97}, as well as landmarking \citep{van07}}. 
We conclude the section with some remarks on the interpretation of multi-state models for reserving. As we explain in the next paragraphs, this is a fundamental step to construct our reserving model.

\subsection{The probabilistic setup}

Let $J=\left(J_z\right)_{z \geqslant 0}$ be a non-explosive pure jump process on a finite state space $\mathcal{S}$, and take $(\Omega, \mathcal{F}, \amsmathbb{P})$ to be the underlying probability space. Here, we let $\mathcal{S}=\{1, \ldots, k\}$, with $k\in \amsmathbb{N}$. Denote by $Y$ the possibly infinite absorption time of $J$. 
Let $X$ be a random variable with values in $\amsmathbb{R}^d$ equipped with the Borel $\sigma-$algebra. We assume that the distribution of $X$ admits a density $g$ with respect to the Lebesgue measure $\lambda$. {\color{black} Absolute continuity in case of known atoms may be relaxed as discussed in \citet{bladt23}.}

We introduce a multivariate counting process $N$ with components $N_{j h}=\left(N_{j h}(z)\right)_{z \geqslant 0}$ given by 
$$N_{j h}(z)=\#\left\{s \in(0, z]: J_{s-}=j, J_s=h\right\}, \quad z \geqslant 0,$$
for $j, h \in \mathcal{S}, j \neq h$. Let us introduce the following assumption.

\begin{assumption}
\label{assumption1}
For every $z\ge 0$ and $j,h\in\mathcal{S},j\neq h$ there exists some $\delta > 0$ such that
$$\amsmathbb{E}[N_{jh}(z)^{1+\delta}]<\infty.$$
\end{assumption} 

{\color{black} Taking the largest $z$ and the minimum across the whole states provides a delta which is independent of $j,h,z$.}

For any $x$ in the support of $X$ define the conditional occupation probabilities according to
$$p_j(z \mid x)=\amsmathbb{E}\left[\mathbbm{1}_{\left\{J_z=j\right\}} \mid X=x\right] .$$
We denote by $p(z \mid x)$ the row vector with elements $p_j(z \mid x)$. We now introduce the cumulative conditional transition rates and cast the conditional occupation probabilities as a product integral of these cumulative rates. To this end, let
$$p_{j h}(z \mid x)=\amsmathbb{E}\left[N_{j h}(z) \mid X=x\right],$$
and define the cumulative conditional transition rates via

{
\color{black}
$$\Lambda_{j h}(z \mid x)=\int_0^z \frac{1}{p_j(s-\mid x)} p_{j h}(\mathrm{~d} s \mid x).$$

A cumulative transition hazard matrix, is defined by having the $j,j$ entry $\Lambda_{j,j}(z \mid x)=\sum_{j\neq}\Lambda_{j,h}(z \mid x)$ for $j, h \in \mathcal{S}, j \neq h$.

}

Let $q(\cdot \mid x)$ be the interval function associated with the conditional transition probabilities:

{\color{black}
$$q_{j h}(s, z \mid x)=\mathbbm{1}_{\left\{p_j(s|x)>0\right\}} \frac{\amsmathbb{E}\left[\mathbbm{1}_{\left\{J_z=h\right\}} \mathbbm{1}_{\left\{J_s=j\right\}} \mid X=x\right]}{p_j(s|x)}+\mathbbm{1}_{\left\{p_j(s|x)=0\right\}} \mathbbm{1}_{\{j=h\}}.$$}

In \citet{overgaard19}, it is shown that this assumption ensures that the product integral 

$$
\Prodi_0^t(\operatorname{Id}+\Lambda(\mathrm{d} s \mid x))
$$

is well-defined. Therefore,

\begin{equation}
    \label{eq:occprob}
    p(t \mid x)=p(0 \mid x) \Prodi_0^t(\operatorname{Id}+\Lambda(\mathrm{d} s \mid x)).
\end{equation}


We are interested in the situation where the observation of $J$ is right-censored. To this end we introduce a strictly positive random variable $W$ describing right-censoring, so that we actually observe the triplet
$$\left(X,\left(J_z\right)_{0 \leqslant z \leqslant W}, Y \wedge W \right) .$$
 Whenever the Markov property fails for the jump process $J$, the following assumption is key for our estimation procedure.

\begin{assumption} Right-censoring is conditionally entirely random:

$$ W \indep J  \mid X.$$

Other censoring schemes are possible if $J$ is Markovian, though we do not make that assumption in the sequel.
\end{assumption}

\begin{remark}
From a reserving perspective, the above assumption amounts to stating that, conditional on covariates, closure of a claim is simply specified through an independent random mechanism. One should note that this assumption is not verifiable, despite its widespread use in survival analysis. 

\end{remark}

We introduce

$$p_j^c(z \mid x)  =\amsmathbb{E}\left[\mathbbm{1}_{\left\{J_z=j\right\}} \mathbbm{1}_{\{z<W\}} \mid X=x\right],$$

and

$$p_{j h}^c(z \mid x)  =\amsmathbb{E}\left[N_{j h}(z \wedge W) \mid X=x\right] .$$

from which we may conclude that

{
\color{black}
\begin{equation}
\label{eq:cumulativetransitionintensities}
\Lambda_{j h}(z \mid x)=\int_0^z \frac{1}{p_j^c(s-\mid x)} p_{j h}^c(\mathrm{~d} s \mid x),
\end{equation}
}
within the interior of the support of $W$.

\subsubsection{Estimators}

Consider the i.i.d. replicates $\left(X^{i},\left(J_z^{i}\right)_{0 \leqslant z \leqslant W^{i}}, Y^{i} \wedge W^{i}\right)_{i=1}^n$. Define {\color{black}$\delta^i:=\mathbbm{1}_{\left\{Y^i\le W^i\right\}}$}, which equals $1$ when the observation is absorbed (closed claim), and zero otherwise. In a reserving context, this distinction is usually reflected by specifying two types of claims: closed claims and RBNS (open) claims, and we write respectively $n=n^{\texttt{Closed}}+n^{\texttt{RBNS}}$.

We now construct an estimator for the conditional occupation probabilities $p_j(z\mid x)$, but from \Cref{eq:occprob}, it essentially suffices to estimate the cumulative conditional transition rates $\Lambda_{j h}(z \mid x)$, which we do using \Cref{eq:cumulativetransitionintensities}.
Let $K_b$ be kernel functions of bounded variation and bounded support, with {\color{black} $b=1,\ldots,d$} and let $(a_n)$ be a bandwidth sequence. We impose the following standard assumption, confering with \citet{winfried86}.

\begin{assumption} The band sequence satisfies:

\begin{enumerate}
    \item $a_n \rightarrow 0$ as $n \rightarrow \infty$.
    \item $n a_n^d \rightarrow \infty$ as $n \rightarrow \infty$.
    \item $\sum_{n=1}^{\infty} c_n^r<\infty$ for some $r \in(1,1+\delta)$ where $c_n:=\left(n a_n^d\right)^{-1} \log n$.
\end{enumerate}

\end{assumption}

{\color{black} For instance, one may take, for $(1+\delta)^{-1}<\eta<1$, any sequence satisfying }

$$a_n^d \sim \frac{\log n}{n^{1-\eta}}.$$

We introduce the usual kernel estimator for the density $g$:

$$\mathbbm{g}^{(n)}(x)=\frac{1}{n} \sum_{i=1}^n \mathbbm{g}^{(n, i)}(x)=\frac{1}{n} \sum_{i=1}^n \prod_{b=1}^d \frac{1}{a_n} K_b\left(\frac{x_b-X_b^{i}}{a_n}\right) .$$

{\color{black} It is well-known (cf. \cite[Ch.~1.4]{KS}) that }

$$\mathbbm{g}^{(n)}(x) \stackrel{\text { a.s. }}{\rightarrow} g(x), \quad n \rightarrow \infty,$$

and, for any random variable $V$ with i.i.d. replicates $(V^i)$ satisfying $\amsmathbb{E}[|V|^{1+\delta}]<\infty$ that

\begin{align}
\label{eq:convas}
\frac{1}{n} \frac{\sum_{i=1}^n V^{i} \mathbbm{g}^{(n, i)}(x)}{\mathbbm{g}^{(n)}(x)} \stackrel{\operatorname{a.s.}}{\rightarrow} \amsmathbb{E}[V \mid X=x], \quad n \rightarrow \infty .
\end{align}

Based on the kernel estimator for $g$, we form the following Nadaraya–Watson type kernel estimators:

\begin{align}
& \amsmathbb{N}_{j h}^{(n)}(z \mid x)=\frac{1}{n} \sum_{i=1}^n \amsmathbb{N}_{j h}^{(n, i)}(z \mid x)=\frac{1}{n} \sum_{i=1}^n \frac{N_{j h}^{i}\left(z \wedge W^i\right) \mathbbm{g}^{(n, i)}(x)}{\mathbbm{g}^{(n)}(x)} \\
& \amsmathbb{I}_j^{(n)}(z \mid x)=\frac{1}{n} \sum_{i=1}^n \amsmathbb{I}_j^{(n, i)}(z \mid x)=\frac{1}{n} \sum_{i=1}^n \frac{\mathbbm{1}_{\left\{z<W^{i}\right\}} \mathbbm{1}_{\left\{J_z^{i}=j\right\} }\mathbbm{g}^{(n, i)}(x)}{\mathbbm{g}^{(n)}(x)} \\
&
\end{align}

\begin{remark} 
Due to \Cref{assumption1}, it follows from \Cref{eq:convas} that 

$$
\amsmathbb{N}_{j h}^{(n)}(z \mid x) \stackrel{\text { a.s. }}{\rightarrow} p_{j h}^c(z \mid x), \quad n \rightarrow \infty.
$$

Furthermore, it holds that

$$
\amsmathbb{I}_j^{(n)}(z \mid x) \stackrel{\text { a.s. }}{\rightarrow} p_j^c(z \mid x), \quad n \rightarrow \infty .
$$

\end{remark}

\begin{definition}
The conditional Nelson–Aalen estimator for the cumulative conditional transition rates is defined as

$$
\mathbb{\Lambda}_{j h}^{(n)}(z \mid x)=\int_0^z \frac{1}{\amsmathbb{I}_j^{(n)}(s-\mid x)} \amsmathbb{N}_{j h}^{(n)}(\mathrm{d} s \mid x) .
$$

\end{definition}

\begin{definition}
The conditional Aalen–Johansen estimator for the conditional occupation probabilities is defined according \Cref{eq:occprob}, but with the cumulative conditional transition
rates replaced by the conditional Nelson–Aalen estimator, that is

$$
\mathbb{p}^{(n)}(z \mid x)=\mathbb{p}^{(n)}(0 \mid x) \Prodi_0^z\left(\operatorname{Id}+\mathbb{\Lambda}^{(n)}(\mathrm{d} s \mid x)\right),
$$

where $\mathbb{p}_j^{(n)}(0 \mid x)=\amsmathbb{I}_j^{(n)}(0 \mid x).$ {\color{black} The estimator $\mathbb{p}_j^{(n)}(0 \mid x)$ for the occupation probabilities $p_j(z|x)$ denotes the j-th entry of the row vector $\mathbb{p}^{(n)}(z \mid x)$, an estimator for $p(z|x)$.}

\begin{remark}
We refer to \citet{bladt23} for the proof that the estimators are pathwise strongly consistent and asymptotically Gaussian distributed.
\end{remark}

\end{definition}

When all the subjects start in state $1$ at inception, we can use the estimator of the occupation probabilities of the absorbing state $k$, $\mathbb{p}_k^{(n)}$ to model the cumulative density function of the individual claim size.

\begin{remark}
The Volterra integral equation defines the product integral of the matrix function $\Lambda(\mathrm{d} s \mid x)$ as the unique solution $p(\cdot \mid x)$ to the equation
        \begin{align*}
           p(t \mid x)=\operatorname{Id}+\int_{(0,t]}p(s-\mid x) \Lambda(\mathrm{d} s \mid x).
        \end{align*}
        
    However, when a piecewise constant estimator of $\Lambda$ is given by $\mathbb{\Lambda}^{(n)}$, as is described in detail in the previous section, the computation of the product integral simplifies. More precisely, denote $0=t_0<t_1<t_2<\cdots$ the sequence of time points  describing all observed jump times in the sample. Then the recursion
\begin{align*}
\mathbb{p}^{(n)}_j(t_{\ell+1} |x)-\mathbb{p}^{(n)}_j(t_{\ell} |x)=\sum_{\substack{k \in \mathcal{Z} \\ k \neq j}} \mathbb{p}^{(n)}_k(t_{\ell} | x)\Delta\mathbb{\Lambda}_{kj}^{(n)}(t_{\ell +1} | x) -\mathbb{p}^{(n)}_j(t_{\ell} | x) \sum_{\substack{k \in \mathcal{Z} \\ k \neq j}} \Delta\mathbb{\Lambda}_{jk}^{(n)}(t_{\ell +1} | x)
\end{align*}
and starting value $\mathbb{p}^{(n)}(0 | x) = \amsmathbb{I}^{(n)}(0 | x)$ completely characterize the product integral $\mathbb{p}^{(n)}_j(\cdot |x)$ at jump times (and constant otherwise).
\end{remark}

\subsection{Practical considerations for claims development}

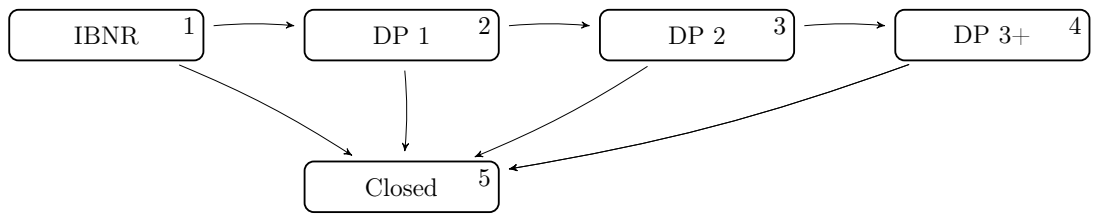
\begin{figure}[!htbp]
	\centering
	\scalebox{0.88}{
	\begin{tikzpicture}[node distance=3em and 0em]
 
		\node[punktl, draw = none] (0) {};
  
		\node[punktl, left = 15mm of 0] (1) {IBNR};
		\node[anchor=north east, at=(1.north east)]{$1$};

            \node[punktl, right = 15mm of 1] (2) {DP 1};
		\node[anchor=north east, at=(2.north east)]{$2$};

            \node[punktl, right = 15mm of 2] (3) {DP 2};
		\node[anchor=north east, at=(3.north east)]{$3$};

            \node[punktl, right = 15mm of 3] (4) {DP 3+};
		\node[anchor=north east, at=(4.north east)]{$4$};
  
		\node[punktl, below = 15mm of 2] (5) {Closed};
		\node[anchor=north east, at=(5.north east)]{$5$};
  


	\path
		(1)	edge [pil, bend left = 5]		node [below]		{}				(0)
  (1)	edge [pil, bend left = 5]		node [below]		{}				(5)
  (2)	edge [pil, bend left = 5]		node [below]		{}				(3)
    (2)	edge [pil, bend left = 5]		node [below]		{}				(5)
    (3)	edge [pil, bend left = 5]		node [below]		{}				(4)
    (3)	edge [pil, bend left = 5]		node [below]		{}				(5)
    (4)	edge [pil, bend left = 5]		node [below]		{}				(5)
  (4)	edge [pil, bend left = 5]		node [below]		{}				(5)

	;
	\end{tikzpicture}}
	\caption{\label{fig:msm} Example of multi-state model for claims reserving, with $k=5$. We interpret time spent in each state as increasing claims size clock, instead of calendar time. The states represent the development periods (DP's) of the individual claims}.
	
\end{figure}

In this section, we offer practical insights into our modeling approach, with the aim of assisting the reader in understanding our multi-state setup for claims reserving. Refer to \Cref{fig:msm}, which conceptually illustrates our model featuring a state space $\mathcal{S}$ with $k=5$ states.

In our framework, we represent the active state as IBNR claims, while we treat RBNS claims as transitioning through intermediate states. Any claim that reaches state $k$ is considered closed in our system due to the presence of an absorbing state. Consequently, we do not allow claims to reopen in our framework. Notably, unlike a traditional survival study, we do not explicitly model the time component, and the number of IBNR claims, along with their characteristics, remains unknown at the time of evaluation. In \Cref{ss:ibnrrbns}, we  delve into how to model RBNS and IBNR claims, showcasing how our model can accommodate such scenarios. Furthermore, we have designed our state space so that state DP3+ at step $k-1$ in \Cref{fig:msm} is the final relevant DP that we explicitly model.

Finally, it is worth mentioning that we have not introduced development triangles yet, which are the standard aggregate datasets commonly used in claims reserving applications. The choice of parameter $k$ aligns with the data that, in a conventional modeling approach, would be aggregated into development triangles of varying sizes. Since our proposed framework does not explicitly model the time component, we assume that payments close at the development year following the last payment. This assumption results in a development triangle with $k-1$ development times. This aspect will be particularly relevant for our data application, where we compare our model to the benchmark chain ladder method outlined in \citet{mack93}, and we need to construct a development triangle for this purpose.

\begin{remark}
    By changing the multi-state model in \Cref{fig:msm}, our approach can be modified to take into account additional domain-specific aspects of reserving data. For instance, if we allow transitions back from the state $k$ (closed) to states $3, \dots, k-2$ (DP $2$, $\dots$), our framework can be modified to take into account reopenings. 
\end{remark}

\section{Claims reserving}
\label{ss:ibnrrbns}

In this section, we will introduce the definitions of final cost of the claims and of the claims reserve, and provide a notation for development triangles, an aggregated representation of the reserving data. We then connect the multi-state model that we introduced in the previous section to the actuarial literature in reserving, with a comparative literature review that distinguishes between the approaches that rely on aggregated data and those that rely on individual data, as the methodology outlined in this manuscript. We remark that aggregated data are a known data representation in survival studies when the individual data are not available, see \citet[p.~168]{andersen12} and \citet{gill86}. We conclude this section with a discussion on the numerical estimation of the total final cost of the claims using the conditional Aalen--Johansen estimator $\mathbb{p}_k^{(n)}$ of $p_k(z|x)$. 

\textbf{Notation.} Together with our sample, which is composed of $n$ i.i.d. observations generated from $J$ and its covariates (refer to the previous section), we additionally observe the following quantities that are \textit{not} related to $J^1, \ldots, J^n$:
{\color{black}$$\left\{T^i, U^i \right\}_{i=1}^n.$$}
In detail, {\color{black} $U^i$ and $T^i$} represent the accident period and the reporting delay of the $i$-th claim, with {\color{black} $U^i, T^i \in \left\{1, \ldots, k-1 \right\}$}. Let us also define the development triangle of the reporting delays as the set 
$$
\mathcal{D}^{(k-1)} = \left\{ d_{\ell,j} : \ell + j \leq k; \ell,j = 1, \ldots, k-1 \right\},
$$
with {\color{black} $ d_{\ell,j} = \sum^{n}_{i=1} \mathbbm{1}_{\left\{U^i = \ell \texttt{ and }  T^i = j\right\}}$} denoting the total claims reported in accident period $\ell$ with delay $j$. 
In a similar fashion, we can define the development triangle of cumulative payments,

$$
\mathcal{C}^{(k-1)} = \left\{ C_{\ell,j} : \ell + j \leq k; \ell,j = 1, \ldots, k-1 \right\},
$$

where {\color{black} $C_{\ell,j}= \sum^n_{i=1}\mathbbm{1}_{\left\{U^i = \ell \texttt{ and }  J^i_{z_{i,j}} = j\right\}} z_{i,j}$}, with $0\leq z_{i,j}\leq W^{i}$ such that  $J^i_{z^-_{i,j}}\in \left\{1, \ldots, j-1\right\}, J^i_{z_{i,j}}=j $ denoting the total cumulative amount paid in accident period $\ell$ and DP $j$.

We are presently interested forecasting the ultimate cost of our claims, {\color{black}$$Y^{\texttt{Closed}}+Y^{\texttt{RBNS}}:=\sum_{i=1}^n \delta^i Y^i+\sum_{i=1}^n (1-\delta^i) Y^i=\sum^{k-1}_{\ell=1} C_{\ell,k-1},$$}
The claims reserve is simply {\color{black}$R= \sum^{k-1}_{\ell=1} \left(C_{\ell,k-1}- C_{\ell,k-\ell}\right)$}.

\subsection{A comparative literature review}

Research papers on loss reserving can be broadly categorized into two main streams of research: models based on development triangles (aggregate loss reserving models), and models based on individual data (individual loss reserving models). The relationship between individual data and aggregate modeling has been extensively studied in \citet{miranda12}, \citet{hiabu17}, and \citet{bischofberger20}, utilizing different survival analysis tools.

The existing body of literature on aggregate reserving has already provided the building blocks for predicting the distribution of loss reserves, with a specific emphasis on models that seek to emulate the empirical chain-ladder algorithm. In this section, we briefly mention some of these models and defer to the comprehensive discussion in \citet{hess02} for their detailed categorization. For a distribution-independent approach to computing the mean squared prediction error of the loss reserve, one can refer to \citet{mack93}, where the author decomposes the prediction error of the loss reserve into its elemental components, namely process variance and estimation variance, as expressed by $\amsmathbb{E}[(\hat R- R)^2]=\Var(R)+\amsmathbb{E}[(\hat R- \amsmathbb{E} [\hat R])^2]$. The model has received several statistical critiques through the years but remains one of the best-performing reserving models in an actuary's toolkit.

In \citet{england99}, the authors replicate the point estimates of the chain-ladder method and ingeniously employ bootstrap and generalized linear models (GLM) to compute $\amsmathbb{E}[(\hat R- R)^2]$. For an extension of the model, allowing for negative payments as well as an exploration of the model assumptions underpinning it, one may consult \citet{verrall00}. An interesting comparison of the fundamental assumptions underpinning the aforementioned models is presented in \citet{mack00}, where the authors conclude that only the approach outlined in \citet{mack93} is the true underlying model of the chain-ladder algorithm.
In a more recent development, in \citet{almudafer22}, the approach in \citet{england99} is integrated into the framework of mixture density networks \citep{bishop94}.

Shifting our focus to non-parametric models for individual reserving, the current discussion primarily centers on providing precise point forecasts for the loss reserve. In contrast, \citet{delong23} discusses the role of the variance function in regression problems for pricing.

In another recent development, \citet{wuthrich18} incorporates neural networks into the chain-ladder model, introducing features to the existing methodology. Further related work can be found in \citet{lopez2019a} and \citet{lopez21}, where the authors employ a tree-based algorithm for modeling RBNS claims using censored data. In particular, in \ \citet{lopez21}, the use of bootstrapping for estimating the uncertainty in the claims reserve is discussed.

Our manuscript introduces a novel approach to the individual reserving problem, allowing us to explicitly model curves of individual claim sizes denoted as $p_k(z|x)$ within our framework.

\subsection{Prediction of RBNS and IBNR claim costs}

We can calculate the final cost of RBNS claims numerically using the conditional Aalen--Johansen estimate $\mathbb{p}_k^{(n)}(z \mid x)$ for $p_k(z|x)$. 
The  {\color{black} predictor} of the final total cost of RBNS claims is defined by

{\color{black}\begin{equation*}
\hat Y^{\texttt{RBNS}}=\sum^n_i\mathbbm{1}_{\left\{\delta^i = 0\right\}} \left(W^{i} + \widehat{\amsmathbb{E}[Y| Y>W^{i}, X=x]}\right)
\end{equation*}}

with 

\begin{equation}
\label{eq:ev}
\widehat{\amsmathbb{E}\left[Y| Y>W^{i}, X=x\right]}  = \frac{1}{1-\mathbb{p}_k^{(n)}(W^i \mid x)}\int^{+\infty}_{W^{i}} (1-\mathbb{p}_k^{(n)}(y \mid x)) \mathrm{d} y.
\end{equation}

We remark that the general formula for the $m$-th moment is explicitly given by

$$
\widehat{\amsmathbb{E}\left[Y^m| Y>W^{i}, X=x\right]}  =  \frac{1}{1-\mathbb{p}_k^{(n)}(W^{i} \mid x)} \int_{W^{i}}^{+\infty} m y^{m-1}(1-\mathbb{p}_k^{(n)}(y \mid x)) \mathrm{d} y.
$$

{\color{black}
\begin{remark}
Relatable approaches to the predictor in \Cref{eq:ev} can be found in the survival literature on censored regression. For instance, similar techniques were presented for the estimation of the linear model in \citet{buckley79}, and for the construction of the artificial data points in \citet{heuchenne07}.
\end{remark}}

Strictly speaking, from a reserving standpoint, we also need to propose a strategy for reserving for the (unknown) IBNR claims. {\color{black} We denote the total number of IBNR claims by $n^{\texttt{IBNR}}$, which is a random variable with values in $\amsmathbb{N}_0$.} The individual cost of the IBNR claims is described by a sequence of iid copies of the random variable $\tilde Y$, namely $\left\{ \tilde Y^{i^\prime} \right\}_{i^\prime \in \amsmathbb{N}}$. Under the framework of \citet{bladt23}, $\tilde Y$ has cumulative distribution function $p_k(z)$. As we will describe shortly, we specify a model that does not consider the features, since in this case we do not have access to future exposures by feature. We propose to correct the model for the random presence of IBNR using a compound distribution, under the assumptions in \citet[p.~148]{klugman98}. We describe the total cost of IBNR claims as follows:

$$Y^{\texttt{IBNR}}=\sum^{n^{\texttt{IBNR}}}_{i^\prime=1} \tilde Y^{i^\prime}.$$
 
The expression for $Y^{\texttt{IBNR}}$ is well known in non-life mathematics and is often referred to as the collective risk model, \citet{parodi14}. The moments of $Y^{\texttt{IBNR}}$ can be calculated in closed form. In particular, the expected cost of IBNR claims is {\color{black} predicted by}

$$\hat Y^{\texttt{IBNR}} = \widehat{\amsmathbb E[Y^{\texttt{IBNR}}]}= \hat n^{\texttt{IBNR}}\widehat{\amsmathbb{E} [\tilde Y]}, $$

with the $m$-th moment of $\tilde Y$ {\color{black} predicted} by

$$
\widehat{ \amsmathbb{E} [\tilde Y^m]} = \int^{\infty}_0 m y^{m-1}(1-\mathbb{p}_k^{(n)}(y)) \mathrm{d} y,
$$

{\color{black} fitting the unconditional Aalen--Johansen to the sample $n$ to obtain $\mathbb{p}_k^{(n)}(z)$. To project the number of IBNR claims, we simply use the chain ladder model on the triangle of reporting delays $\mathcal{D}^{(k-1)}$ and obtain $\hat n^{\texttt{IBNR}}$.}

Similarly, the variance of the IBNR claims is {\color{black} predicted by}
$$\widehat {\mbox{Var}( Y^{\texttt{IBNR}})}= \hat n^{\texttt{IBNR}} \left[\widehat{\amsmathbb{E} [\tilde Y^2]}- \widehat{\amsmathbb{E} [\tilde Y]}^2\right]+\widehat{\amsmathbb{E} [\tilde Y]}^2 \widehat{\mbox{Var}(n^{\texttt{IBNR}})}.$$ 

{\color{black} We denote by $\widehat{\mbox{Var}(n^{\texttt{IBNR}})}$ a predictor for the variance of $n^{\texttt{IBNR}}$. In this manuscript, we consider the predictor for the process variance discussed in \citet{mack93}.}

The total size of claims is $Y^{\texttt{TOT}}= Y^{\texttt{Closed}}+Y^{\texttt{RBNS}}+Y^{\texttt{IBNR}}$ and we estimate it with $\hat Y^{\texttt{TOT}}= Y^{\texttt{Closed}} + \hat Y^{\texttt{RBNS}}+\hat Y^{\texttt{IBNR}}$.
We remark that in several lines of business the contribution of the true cost of IBNR claims ($ Y^{\texttt{IBNR}}$) to $Y^{\texttt{TOT}}$ is minor compared to the cost of RBNS claims ($ Y^{\texttt{RBNS}}$). This is the case for those lines of business that have data with mostly single payments and a short time lag between reporting and settlement. See for instance the results presented in \citet[p.~80]{friedland10} on the US market data, comparing for different lines of insurance the size of the estimated cost of IBNR claims to paid and outstanding claims. As we will show in \Cref{s:realdata}, our real dataset exhibits exactly this behaviour and the IBNR correction will have a minor impact on the estimated final cost.

\section{Simulation of RBNS claims}
\label{s:simulation}

In this section, we perform two simulation studies to provide numerical evidence of the applicability of our models to claims reserving. To demonstrate the potential benefits to actuaries of using our modeling approach, we first compare our results with the chain ladder model \citet{mack93}. {\color{black} A second comparison with} the hierarchical  reserving model in \citet{crevecoeur23} {\color{black} can be found in \Cref{appendix:hirem}}. To facilitate the presentation of our results, we refer to our model as AJ, the chain ladder model in \citet{mack93} as CL, and the hierarchical reserving model as \texttt{hirem}.

\subsection{Implementation}
\label{ss:implementation}

\begin{figure}
    \centering 
\begin{subfigure}[t]{0.4\linewidth}
  \includegraphics[width=\linewidth]{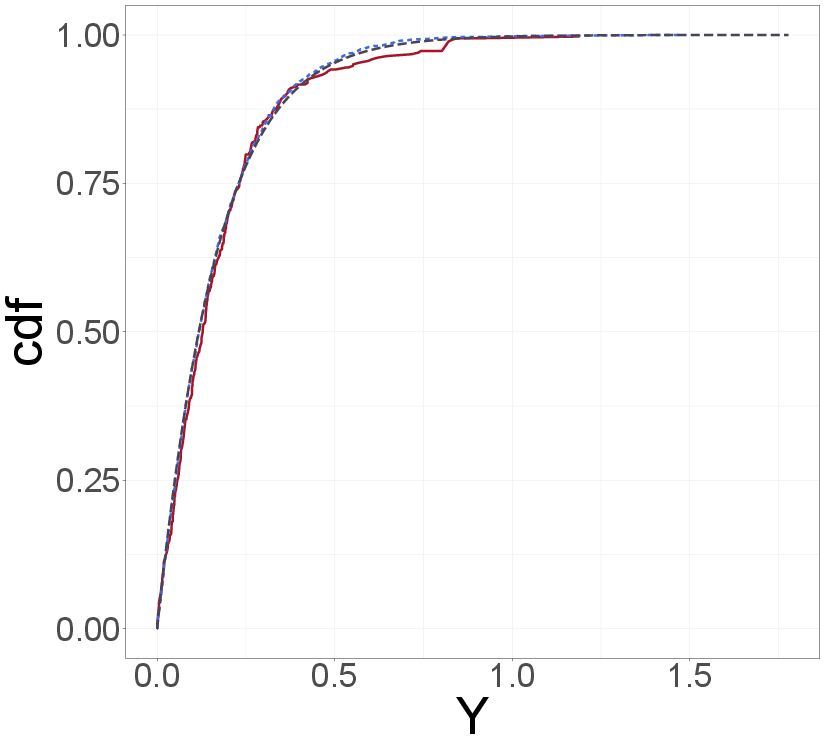}
  \caption{$k=4$.}
  \label{fig:sc4}
\end{subfigure}\hfil
\begin{subfigure}[t]{0.4\linewidth}
  \includegraphics[width=\linewidth]{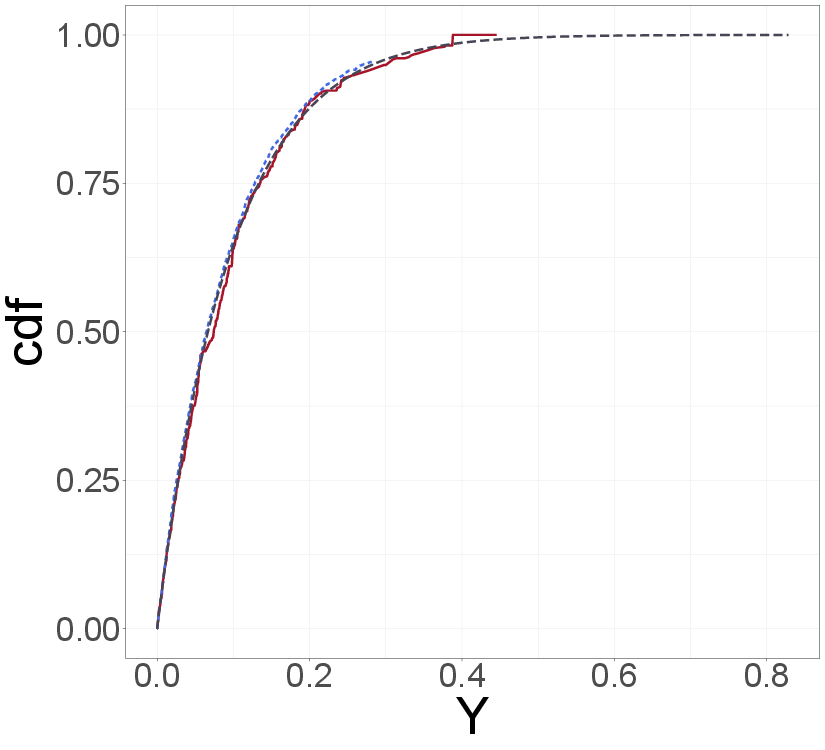}
  \caption{$k=5$.}
  \label{fig:sc5}
\end{subfigure}\hfil
\begin{subfigure}[t]{0.4\linewidth}
  \includegraphics[width=\linewidth]{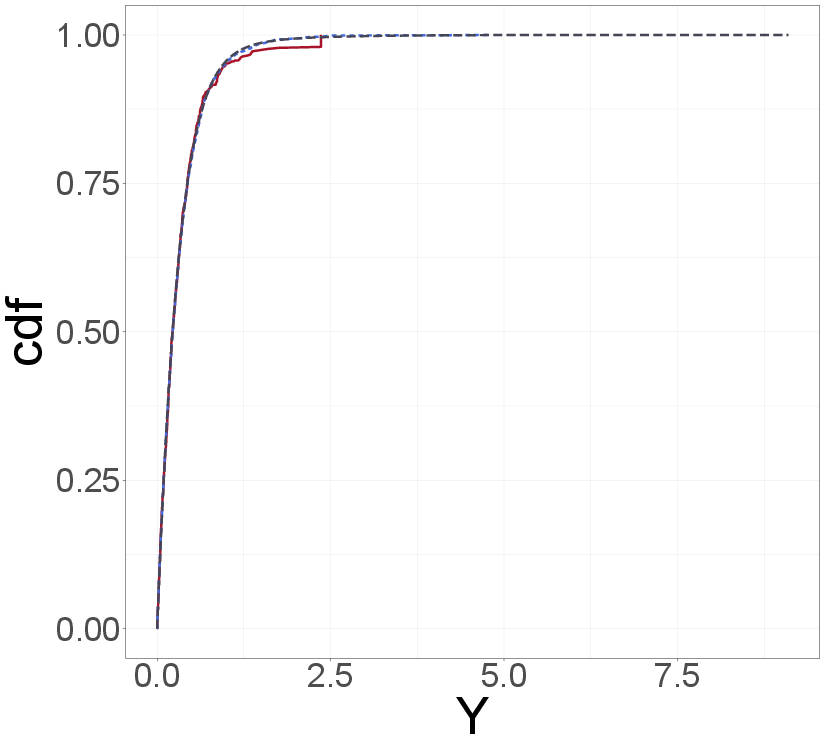}
  \caption{$k=6$.}
  \label{fig:sc6}
\end{subfigure}\hfil
\begin{subfigure}[t]{0.4\linewidth}
  \includegraphics[width=\linewidth]{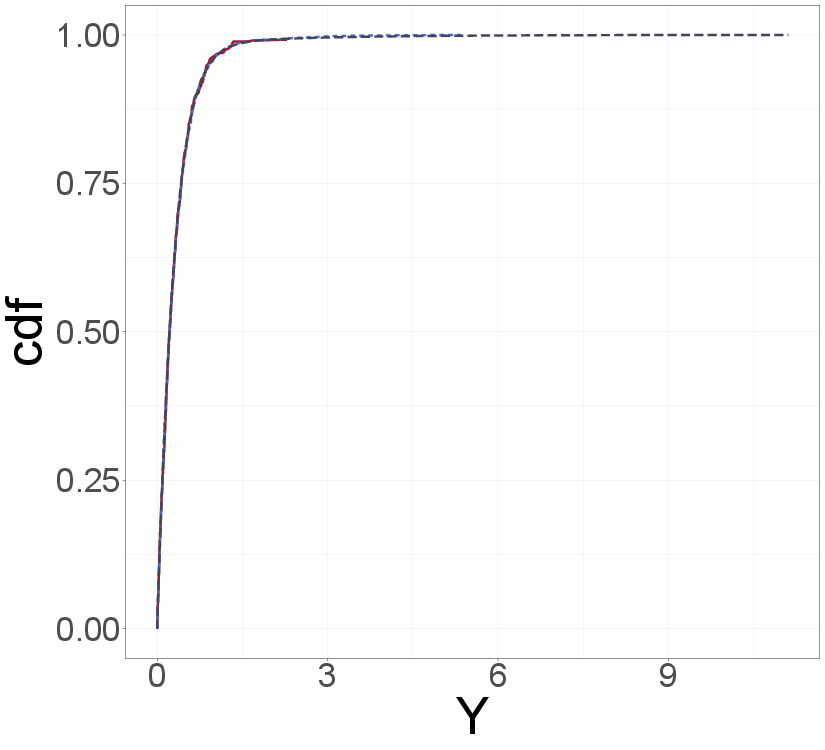}
  \caption{$k=7$.}
  \label{fig:sc7}
\end{subfigure}

\caption{\label{fig:scs} Comparison of the true severity curve (dark grey dotted line) to the fitted severity curve for different data sizes in the 4 simulated scenarios. The red curve is fitted on $120 (k-1)$ RBNS claims, the blue curve is fitted on $1200$ RBNS claims. $Z$ is rounded by millions.}
\end{figure}

It is possible to simulate observations from a jump-process model using the function \texttt{sim\_path} from the \texttt{R} package \texttt{AalenJohansen} \citet{aalenjohanson}, by introducing a matrix of transition intensities. Markov and semi-Markov models are allowed.
To obtain a realistic set of examples for the simulation study, we used the Nelson-Aalen implementation of the \texttt{R} package \texttt{survival} \citet{survivalpackage} to estimate the (unconditional) cumulative transition intensities $\mathbb{\Lambda}_{j h}^{(n)}(z)$ for different choices of $k=4,5,6,7$ on the real dataset available to us for writing this manuscript. Then, similar choices of parameters were employed for the present section. See \Cref{tab:codandata} for the description of the real data. The matrix of intensities can be obtained from the cumulative intensities $\mathbb{\Lambda}_{j h}^{(n)}(z)$.
Simulation by knowing the data generation process underlying the data is important to understand model performance compared to the true estimation target. For example, in \Cref{fig:scs} we show an example of the fitted $\mathbb{p}_k^{(n)}(z \mid x)$ on datasets simulated directly with $\mathbb{\Lambda}$ without any feature effect. For each choice of $k$ we simulated two datasets. The first dataset contains $120 (k-1)$ observations. The second dataset contains $1200 (k-1)$ observations. By comparing the $\mathbb{p}_k^{(n)}(z \mid x)$ fitted to the smallest sample (red) with the curve fitted to the largest sample (blue), we observe that increasing the sample size yields fits that are closer to the true $p_k(z\mid x)$, empirically validating the consistency of the Aalen--Johansen estimator with claim size as operational time.

In particular, we have presented a model that can take into account the individual feature information. In the case studies that we present in the next section, we treat the accident period as a feature and add an effect by simply specifying $q_{j h}(s, z \mid x) = q_{j h}(s, z)/f(x)$, where $f(x)$ is a function of some feature $x$ and we can derive $q_{j h}(s, z)$ from the cumulative transition matrix obtained with the \texttt{survival} package. In the simulations we present, we assume that observations are generated from $f(x)= {10+k-x}$, where in place of $x$ we use the accident period, namely {\color{black}$U^i$} introduced in the previous sections. For the simulation of the reserving data sets, we decide arbitrarily the number of observations in each accident period {\color{black}$U^i$}. Each claimant is then associated with a numeric variable representing the accident period {\color{black}$U^i$}, taking values in $\left\{1 \ldots, k-1\right\}$, from which we can calculate the effect {\color{black}$f(U^i)= {10+k-U^i}$}. We also add an effect on the volumes, and generate fewer observations in most recent accident period. Generating decreasing volumes resembles cycles in the underwriting, a case study of an insurer that underwrites less policies in the more recent accident periods, resulting in less claims. In particular, we generated $1200$ observations in the first accident period {\color{black}$U^i = 1$} and decreased by $100$ the number of observations every accident periods. For instance, we have $1100$ observations with {\color{black}$U^i=2$} and $1000$ observations with {\color{black}$U^i=3$}.

\subsection{Evaluating the models performance}

Using the approach introduced in \Cref{ss:implementation}, we present two simulated scenarios: Alpha and Beta. In scenario Alpha, we simulate directly with unconditional transition intensities that we have fitted to the real data. In the Beta scenario we introduce the effect of the accident period characteristics and the occupation intensities. In each scenario we simulate, for the different choices of $k=4,5,6,7$, a total number of $20$ datasets to smooth possible fluctuations due to the random generator and report the average results that we obtain. Each dataset in scenario Alpha contains $1200(k-1)$ observations. The dataset in scenario Beta contains $1200(k-1)-100(k-2)$ observations. As motivated in \Cref{ss:implementation}, we let the volumes decrease in recent accident periods. We chose arbitrarily to decrease the volume by $100$ in every accident period subsequent to the first one.

\begin{figure}
\centering 
\begin{subfigure}[t]{0.45\linewidth}
  \includegraphics[width=0.75\linewidth]{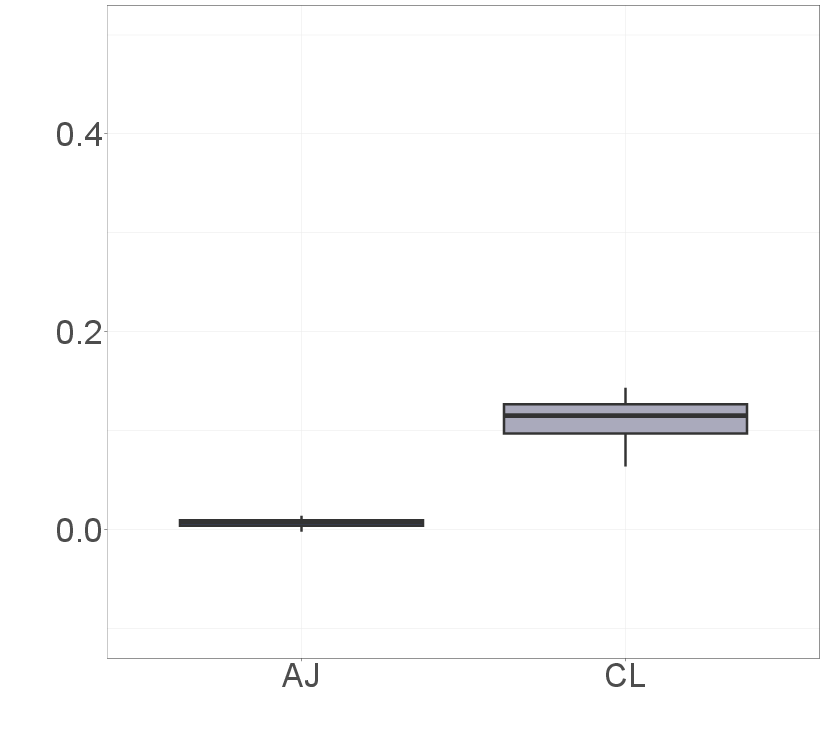}
  \caption{$k=4$, no features.}
  \label{fig:ei_sim_nf_4}
\end{subfigure}\hfil
\begin{subfigure}[t]{0.45\linewidth}
  \includegraphics[width=0.75\linewidth]{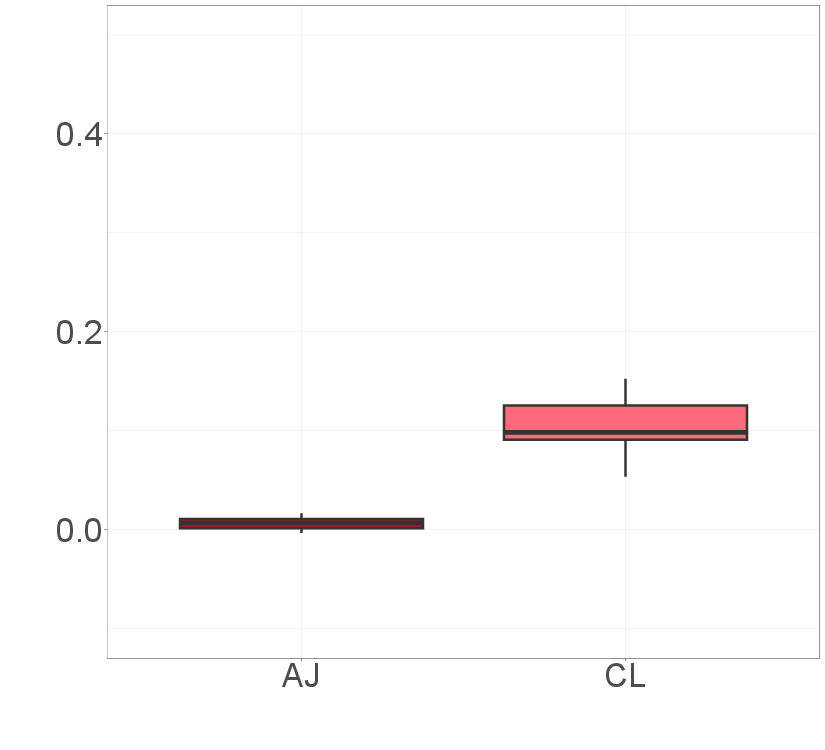}
  \caption{$k=4$, accident period trend.}
  \label{fig:ei_sim_f_4}
\end{subfigure}\hfil
\begin{subfigure}[t]{0.45\linewidth}
  \includegraphics[width=0.75\linewidth]{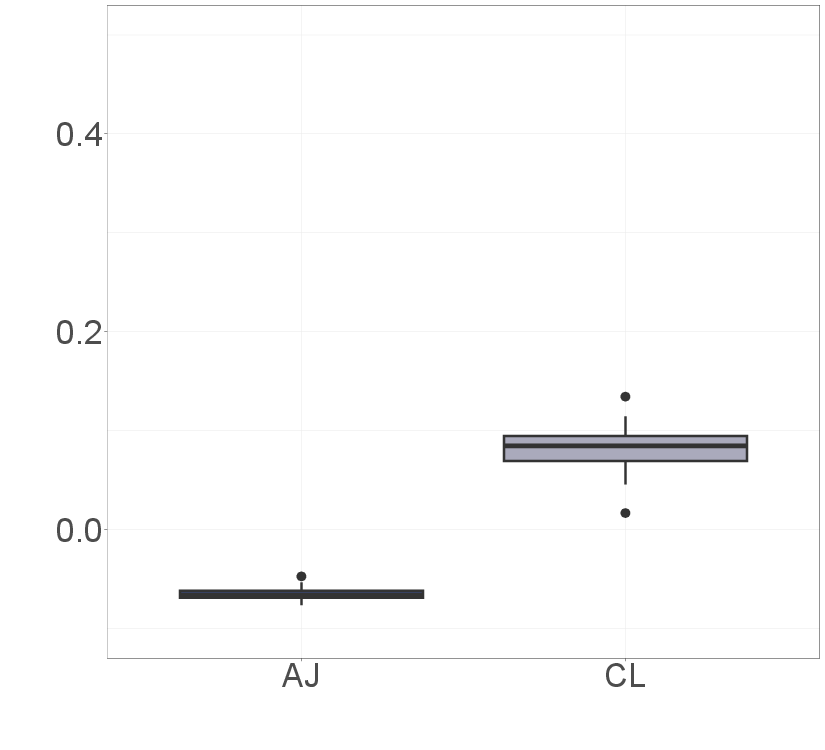}
  \caption{$k=5$, no features.}
  \label{fig:ei_sim_nf_5}
\end{subfigure}\hfil
\begin{subfigure}[t]{0.45\linewidth}
  \includegraphics[width=0.75\linewidth]{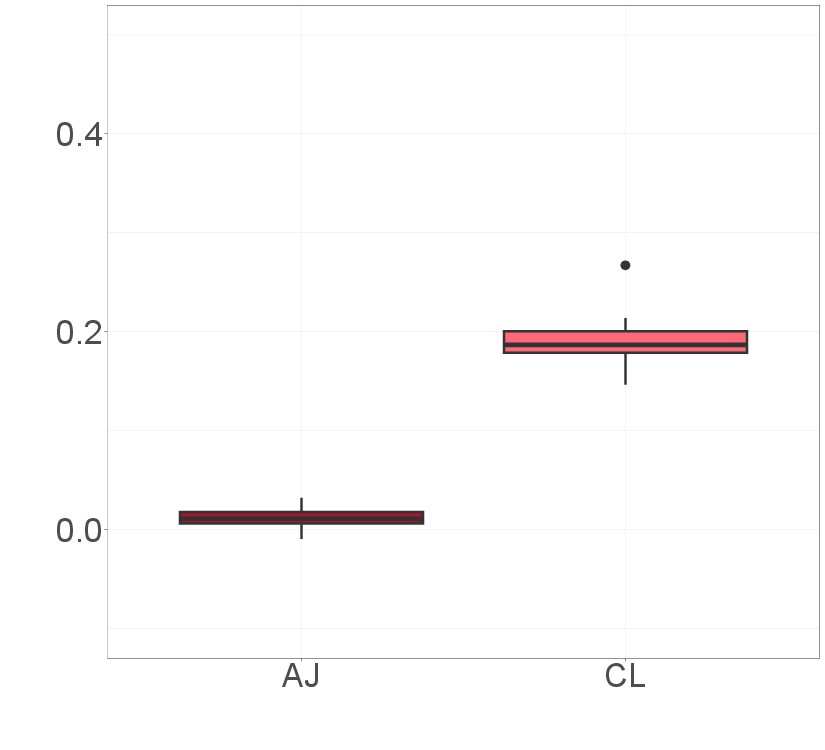}
  \caption{$k=5$, accident period trend.}
  \label{fig:ei_sim_f_5}\end{subfigure}\hfil
\begin{subfigure}[t]{0.45\linewidth}
  \includegraphics[width=0.75\linewidth]{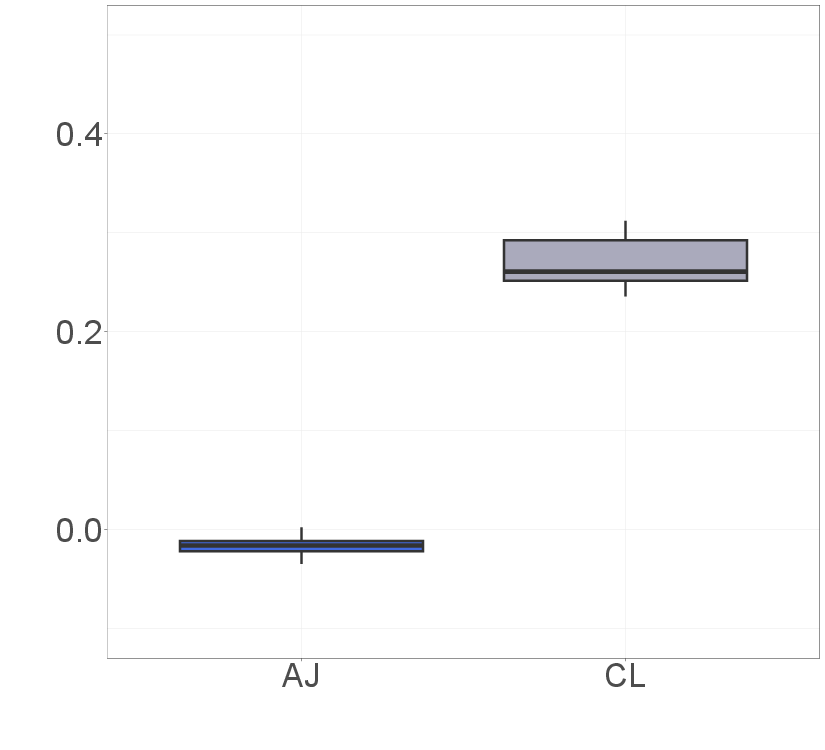}
  \caption{$k=6$, no features.}
  \label{fig:ei_sim_nf_6}
\end{subfigure}\hfil
\begin{subfigure}[t]{0.45\linewidth}
  \includegraphics[width=0.75\linewidth]{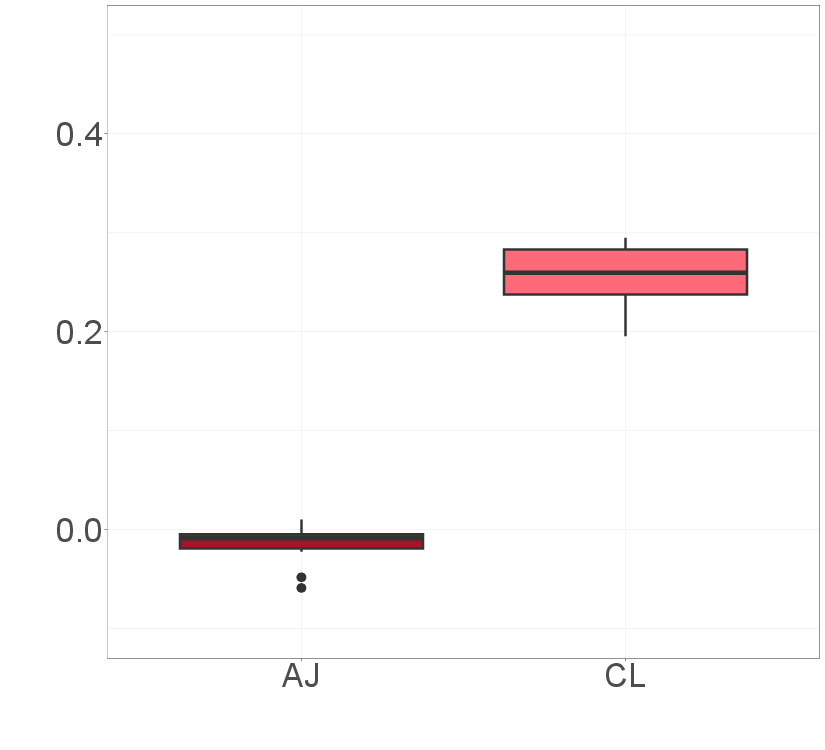}
  \caption{$k=6$, accident period trend.}
  \label{fig:ei_sim_f_6}
\end{subfigure}
\begin{subfigure}[t]{0.45\linewidth}
  \includegraphics[width=0.75\linewidth]{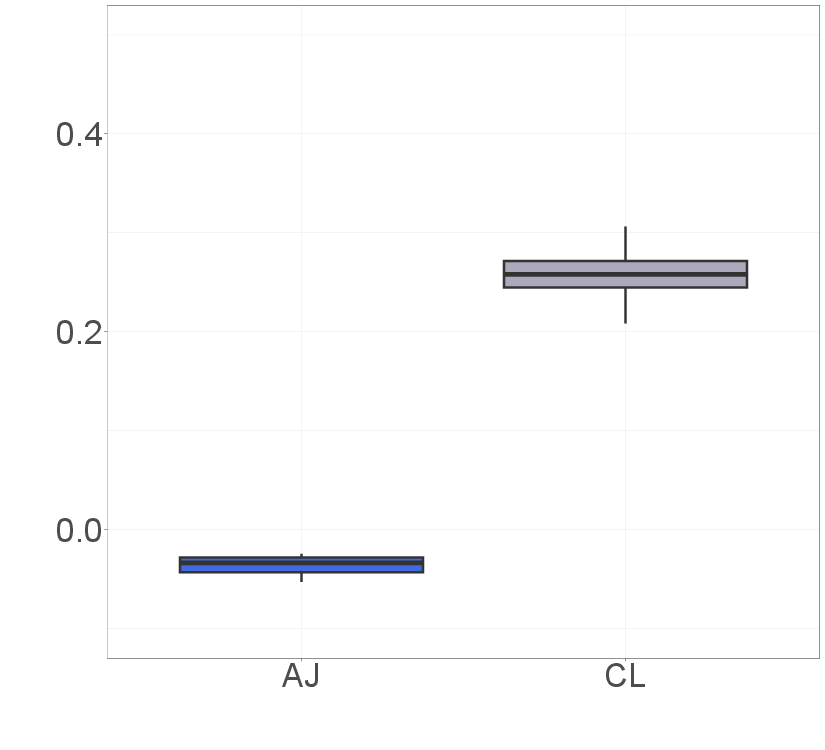}
  \caption{$k=7$, no features.}
  \label{fig:ei_sim_nf_7}
\end{subfigure}\hfil
\begin{subfigure}[t]{0.45\linewidth}
  \includegraphics[width=0.75\linewidth]{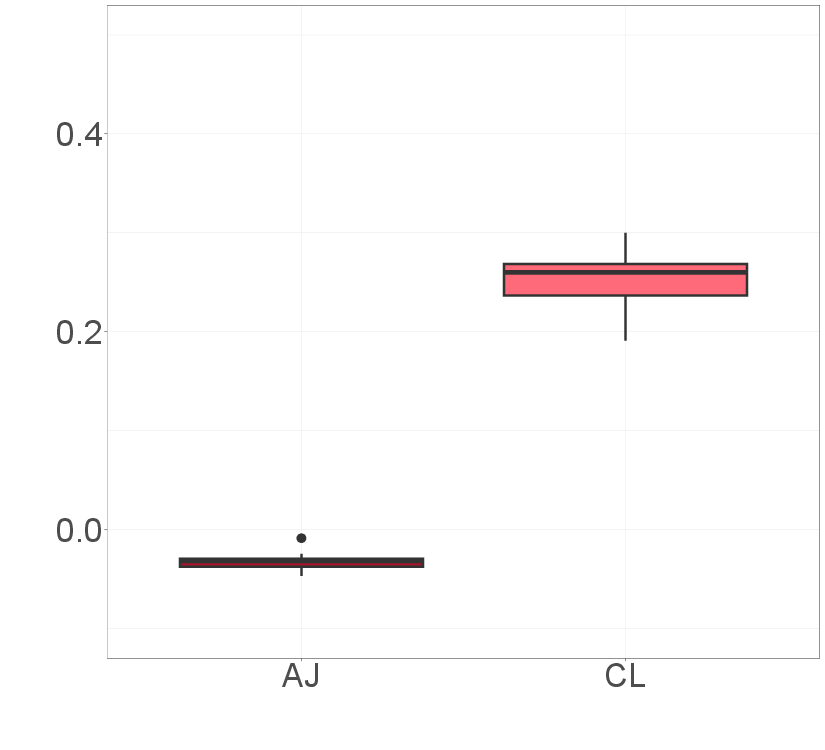}
  \caption{$k=7$, accident period trend.}
  \label{fig:ei_sim_f_7}
\end{subfigure}
\caption{\label{fig:ei_sim} Box plots of \texttt{EI} for AJ and CL over the $20$ simulations, for each value of $k$, in the Alpha scenario (left column) and the Beta scenario (right column). }
\end{figure}

The choice of measure of model performance can be an ardouos task when models are very different, as are AJ and CL. The first performance measure that we propose is the error incidence,
$$\texttt{EI}= \frac{\hat Y^{\texttt{TOT}}}{Y^{\texttt{TOT}}}-1,$$ 
where $\hat Y^{\texttt{TOT}} $ is the predicted final claim size and $Y^{\texttt{TOT}}$ is the actual final claim size, which is available from the simulation. The \texttt{EI} is an interesting benchmark that provides insight into the relative accuracy of the AJ compared to the CL, though it does not highlight the fact that AJ provides a much more nuanced description of the claim size than point estimates. In \Cref{fig:ei_sim} we provide a box plot of the \texttt{EI} for the Alpha scenario (left side) and the Beta scenario (right side) for the different values of $k$. Compared to the CL, the AJ model performs favourably  for each choice of $k$ and in both scenarios. 

While the \texttt{EI} is an interesting benchmark, to better understand the performance of the models compared to the data at hand, we would ideally like to use a proper scoring measure \citet{gneiting07}.  In principle, a scoring metric with the property of being proper is able to evaluate the better model by assigning better scores to better models. In this manuscript we consider the Continuously Ranked Probability Score (CRPS) as a proper scoring metric:
{\color{black}
\begin{align*}
\operatorname{CRPS}(\mathbb{p}_k^{(n)}(z \mid x) ,y) &= 
 \int_{0}^{+\infty} (\mathbb{p}_k^{(n)}(z \mid x) - \mathbbm{1}_{\{y \leq z\}})^2 \mathrm{d} z \\
&= \int^{y}_0 \left(\mathbb{p}_k^{(n)}(z \mid x)\right)^2 \mathrm{d}z + \int^{+\infty}_y \left(1-\mathbb{p}_k^{(n)}(z \mid x)\right)^2 \mathrm{d}z,
\end{align*}}
where $\mathbb{p}_k^{(n)}(z \mid x)$ is the predicted curve of the individual claims sizes and $y_i$ is the (true) value of the final claim size that is available from the simulation. The interested reader can refer to the established statistical literature on proper scoring for a detailed discussion on this topic \citet{gneiting07, gneiting11}.

\subsection{Empirical analysis}
\label{ss:empirical_analysis}

The average results of our $20$ simulations for the Alpha scenario are reported in \Cref{tab:sim_alpha}. Each column of the table represents one of the AJ models fitted for the different choices of $k$ (column one). The target $Y^{\texttt{TOT}}$ (average simulated total cost) assuming $Y^{\texttt{IBNR}}=0$ is in column four. The predicted final cost is $\hat Y^{\texttt{TOT}} = Y^{\texttt{Closed}}+ \hat{Y^{\texttt{RBNS}}}$. The \texttt{EI} results show that, on average, our models outperform the CL for every choice of $k$ except $k=5$ (columns five and six). For all choices of $k$, we find that the relative variability of our model is less than that provided by the CL (columns seven and eight). We compute $\mbox{sd}(Y^{\texttt{TOT}})=\sqrt{\sum^{n^{\texttt{RBNS}}}_i \mbox{Var}(Y^i)}$. In this manuscript we will not discuss the estimation error, and the results we present with respect to the variability of the reserve refer only to the process variance. In column nine we report the (average) CRPS. 

\begin{table}[ht]
\centering
\begin{adjustbox}{max width=\textwidth}
\begin{tabular}{rllcccccc}
  \hline
$k$ & Scenario & $U$ & Actual $Y^{\texttt{TOT}}$ (average) & \texttt{EI} (AJ) & \texttt{EI} (CL) & $\hat{\mbox{sd}}(Y^{\texttt{TOT}})/\hat Y^{\texttt{TOT}}$ (AJ) & $\hat{\mbox{sd}}(Y^{\texttt{TOT}})/\hat Y^{\texttt{TOT}}$ (CL) & CRPS (average) \\ 
  \hline
4 & \multirow{4}{1.1cm}{Alpha} & \xmark & \multirow{1}{1.1cm}{598.483} & 0.006 & 0.111 & 0.008 & 0.009 & 0.145 \\ 

   
  
  5 &  & \xmark & \multirow{1}{1.1cm}{456.746} & -0.085 & 0.082 & 0.007 & 0.008 & 0.095 \\ 
  
   
 
  6 &  & \xmark & \multirow{1}{1.1cm}{1952.683}& -0.016 & 0.269 & 0.007 & 0.009 & 0.410 \\ 
  
   
  
  7&  & \xmark & \multirow{1}{1.1cm}{2359.151} & -0.035 & 0.257 & 0.007 & 0.007 & 0.557 \\
  
   
   \hline
\end{tabular}
\end{adjustbox}
\caption{\label{tab:sim_alpha} Results for scenario Alpha. For each value of $k$ (column one) we present the average results over the $20$ simulations. Each row of the table corresponds to a different AJ model specification (column three). The table includes the (average) actual $Y^{\texttt{TOT}}$ simulated total cost (column four) and the error incidence for the AJ and the CL (columns five and six). In columns seven and eight we reported the coefficients of variation of $Y^{\texttt{TOT}}$. The results for the CRPS are reported in column nine. }
\end{table}

\Cref{tab:sim_beta} reports the average results for the $20$ simulations of the Beta scenario. For each simulation we fit two models, with and without the use of $U$ as a feature (\cmark and \xmark \; in the table). The model with $U$ is correctly specified assuming the Beta scenario. The aim of this experiment is to show that the CRPS is able to judge which is the better model in a context where we know the data generation process. As expected, we obtain a lower average CRPS for the models that include features. The CRPS results in \Cref{tab:sim_beta} are relative to the AJ model with feature $U$. For example, for $k=6$ we see that the AJ model without the feature $U$ shows (on average) a CRPS $3.9 \%$ higher than the CRPS of the AJ model with feature $U$. For the purpose of an easier interpretation, the CRPS results will be shown relative to the AJ with feature model also in the following tables. The (average) target $Y^{\texttt{TOT}}$ is reported in the fourth column. The results for \texttt{EI} indicate that the AJ model generally outperforms the CL model. The relative variability results show comparable results for AJ and CL.  

\begin{table}[ht]
\centering
\begin{adjustbox}{max width=\textwidth}
\begin{tabular}{rllcccccc}
  \hline
$k$ & Scenario & $U$ & Actual $Y^{\texttt{TOT}}$ (average) & \texttt{EI} (AJ) & \texttt{EI} (CL) & $\hat{\mbox{sd}}(Y^{\texttt{TOT}})/\hat Y^{\texttt{TOT}}$ (AJ) &$\hat{\mbox{sd}}(Y^{\texttt{TOT}})/\hat Y^{\texttt{TOT}}$ (CL) & CRPS (average, relative) \\ 
  \hline

   \multirow{2}{.5cm}{4}& \multirow{8}{.5cm}{Beta} & \cmark & \multirow{2}{1.1cm}{42.279} & 0.006 & \multirow{2}{.8cm}{0.103} & 0.008 & \multirow{2}{0.8cm}{0.008} & 1.000 \\ 
   
  &  & \xmark &  & 0.005 &  & 0.008 &  & 1.008 \\ 
  
  
   \multirow{2}{.5cm}{5}& & \cmark & \multirow{2}{1.1cm}{94.580}  & 0.011 & \multirow{2}{.8cm}{0.190} & 0.009 & \multirow{2}{0.8cm}{0.008} & 1.000 \\ 
   
 &  & \xmark &  & -0.006 &  & 0.007 &  & 1.064 \\ 
 
  
   \multirow{2}{.5cm}{6}&  & \cmark & \multirow{2}{1.1cm}{108.423} & -0.013 & \multirow{2}{.8cm}{0.256} & 0.007 & \multirow{2}{0.8cm}{0.008} & 1.000 \\ 
   
  &  & \xmark &  & -0.016 &  & 0.007 &  & 1.039 \\ 
  
  
   \multirow{2}{.5cm}{7}& & \cmark & \multirow{2}{1.1cm}{116.032} & -0.033 & \multirow{2}{.8cm}{0.255} & 0.007 & \multirow{2}{0.8cm}{0.008} & 1.000 \\  
   
   &  & \xmark &  & -0.031 & & 0.008 &  & 1.260 \\ 
   \hline
\end{tabular}
\end{adjustbox}
\caption{\label{tab:sim_beta} Results for scenario Beta. For each value of $k$ (column one) we present the average results over the $20$ simulations. Each row of the table corresponds to a different AJ model specification (column three). The table includes the (average) actual simulated total cost (column four) and the error incidence for the AJ and the CL (columns five and six). In columns seven and eight we reported the coefficients of variation of $Y^{\texttt{TOT}}$. The results for the CRPS, relative to the AJ model with features, are reported in column nine. }
\end{table}

\section{A data application on an insurance portfolio} 
\label{s:realdata}

In this section we propose a case study on our model using real data. For this data application, we have at our disposal a recent real dataset from a Danish non-life insurer. The dataset is not publicly available.  

\begin{table}[ht]
\centering
\begin{tabular}{l|l}
 Covariates & Description\\ \hline
\texttt{Claim\_number} & Policy identifier\\
 \texttt{claim\_type} $\in  \left\{1, \ldots, 20 \right\}$ & Type of claim \\
\texttt{AM} & Accident month\\
\texttt{CM} & Calendar month of report\\
\texttt{DM} & Development month\\
\texttt{incPaid} & Incremental paid amount \\
\texttt{Delta} & Indicator, $0$ when the claim is open \\
 \hline
  \end{tabular}
  \caption{\label{tab:codandata} Description of the real dataset.}
\end{table}

\begin{figure}
    \centering 
\begin{subfigure}[t]{0.4\linewidth}
  \includegraphics[width=\linewidth]{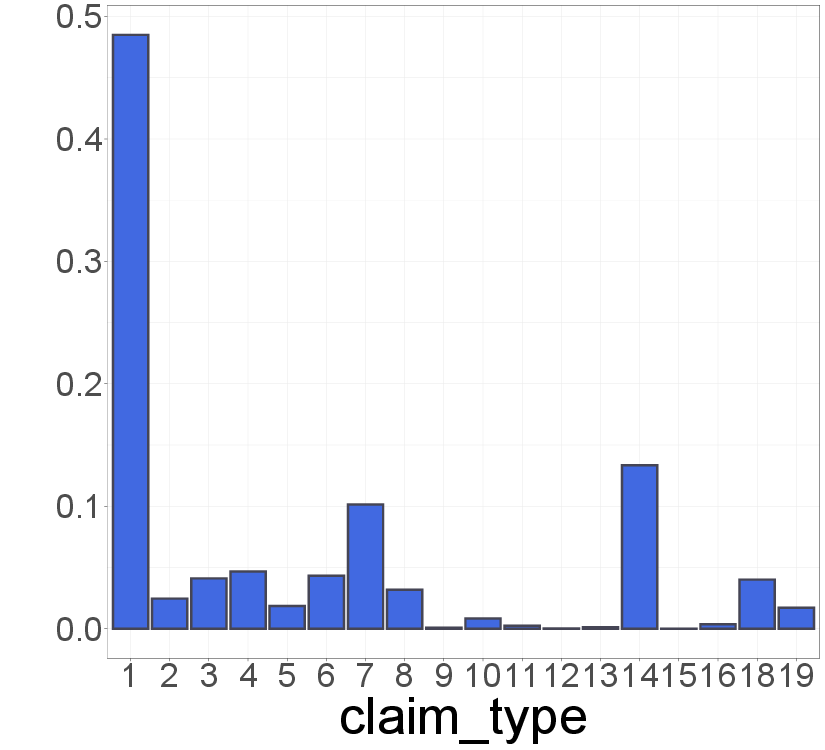}
  \caption{Frequency by \texttt{claim\_type}.}
  \label{fig:eda1}
\end{subfigure}\hfil
\begin{subfigure}[t]{0.4\linewidth}
  \includegraphics[width=\linewidth]{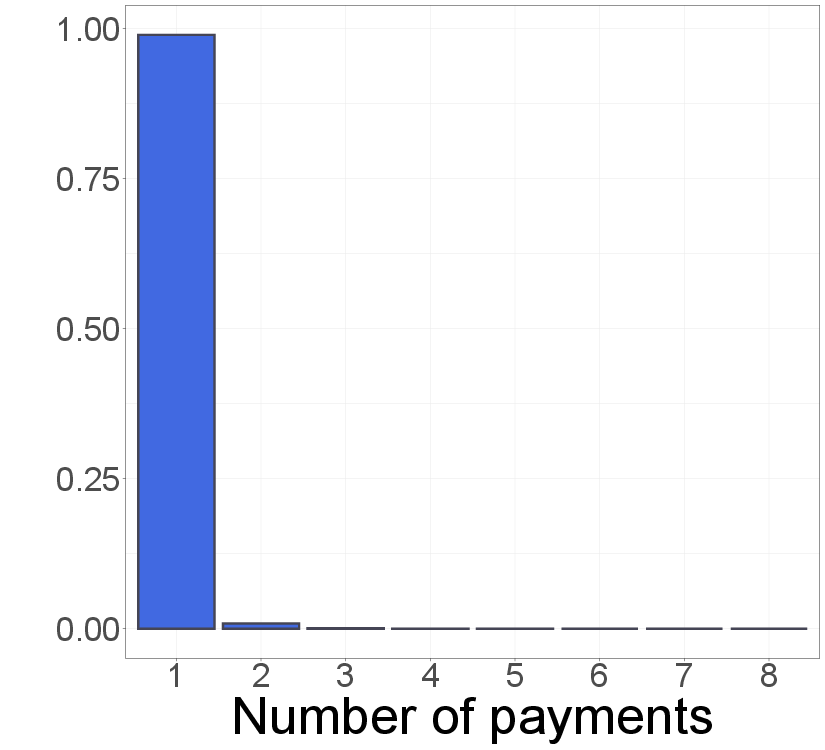}
  \caption{Distribution of the number of payments.}
  \label{fig:eda2}
\end{subfigure}\hfil
\begin{subfigure}[t]{0.4\linewidth}
  \includegraphics[width=\linewidth]{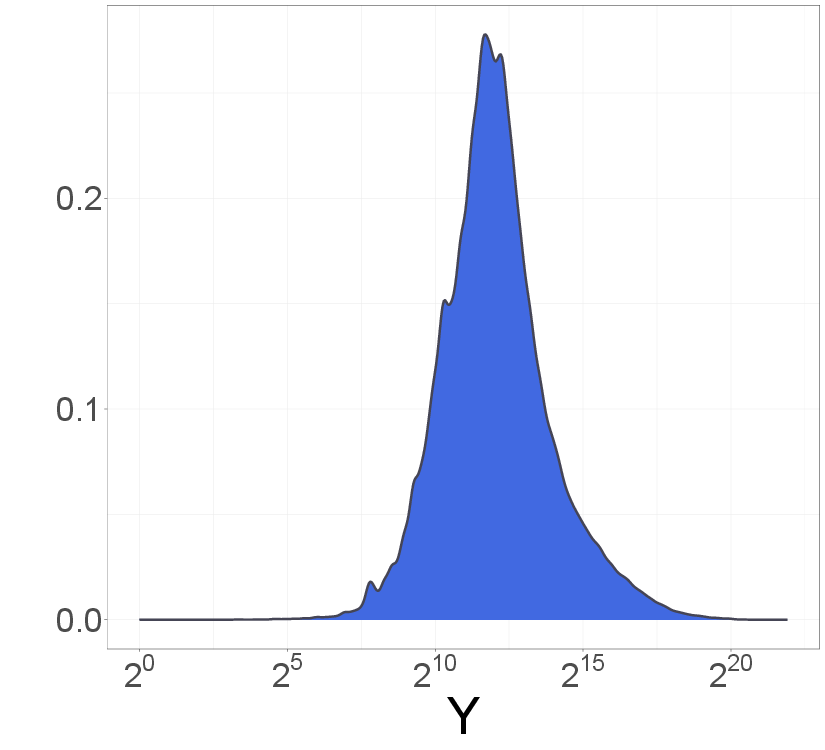}
  \caption{Density plot of the total individual claim size. The x-axis is scaled by  \texttt{log2}.}
  \label{fig:eda3}
\end{subfigure}\hfil
\begin{subfigure}[t]{0.4\linewidth}
  \includegraphics[width=\linewidth]{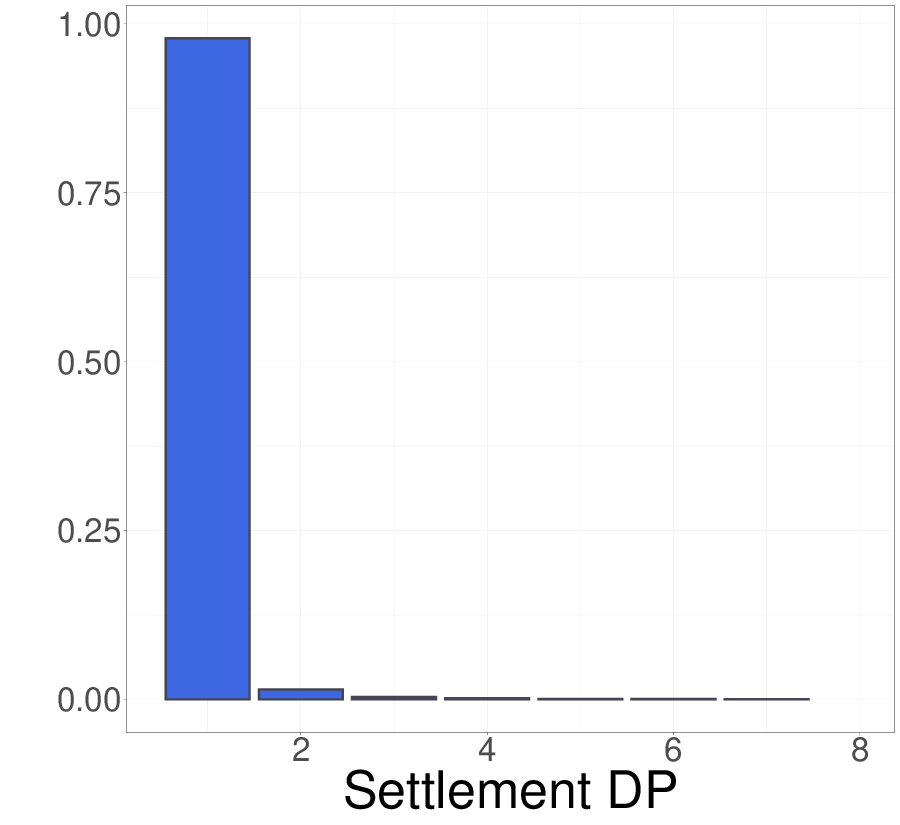}
  \caption{Settlement delay from accident.}
  \label{fig:eda4}
\end{subfigure}

\caption{\label{fig:eda} Exploratory data analysis on the real dataset that we use in this section. We show the relative frequencies by type of claim \Cref{fig:eda1}, the distribution of the number of payments \Cref{fig:eda2}, the density plot of the total  individual claim size \Cref{fig:eda3} and the distribution of the settlement delay from accident \Cref{fig:eda4}. }
\end{figure}

We provide additional insight in \Cref{fig:eda}. The exploratory analysis of the data shows that most of the records have a single payment (\Cref{fig:eda2}) and a very fast settlement (\Cref{fig:eda4}). The data we present in this manuscript comes from a very stable line of business. Interestingly, we show that we are able to outperform the CL on data where the CL is expected to behave as well as it can. Most of the datasets belong to \texttt{claim\_type 1}, see \Cref{fig:eda1}.

\begin{remark}
Notice that the covariate in our application is discrete, which with a uniform kernel effectively amounts to subsampling of data. However, the estimators of the paper provide consistent and asymptotically normal estimators even when covariates are continuous. Nonetheless, kernel-based methods are prone to the curse of dimesionality, and hence our method provides the best results when covariate dimensions are small.
\end{remark}

To analyse the real dataset, we propose two strategies that an actuary could follow to calculate the loss reserve. In \Cref{ss:differentk} we show an example where we censor our data by different calendar periods and fit an AJ model for the maximum depth of data available. For example, for a model with $k=4$ we will have the individual data corresponding to a development triangle with $3$ accident periods. The purpose of this application is to show the behaviour of our model compared to the CL on different datasets. In \Cref{ss:modelselection} we conclude with an application where we highlight the use of the CRPS measure to perform model selection, after having censored the data at an arbitrary calendar time, i.e. $5$ calendar periods. Here we fit different models to the same datasets and select the best performing model in terms of CRPS to calculate the loss reserve. The applications in this section also comprise the IBNR model that we have illustrated in the previous sections. In fact, we predict $\hat Y^{\texttt{TOT}} = Y^{\texttt{Closed}}+ \hat{Y}^{\texttt{RBNS}}+\hat{Y}^{\texttt{IBNR}}$ and $\widehat{\mbox{Var}(Y^{\texttt{TOT}})}=\widehat{\mbox{Var}( Y^{\texttt{RBNS}})}+ \widehat{\mbox{Var}( Y^{\texttt{IBNR}})}$. 

\subsection{Model comparison on different datasets}
\label{ss:differentk}

\begin{figure}
    \centering 
\begin{subfigure}[t]{0.4\linewidth}
  \includegraphics[width=\linewidth]{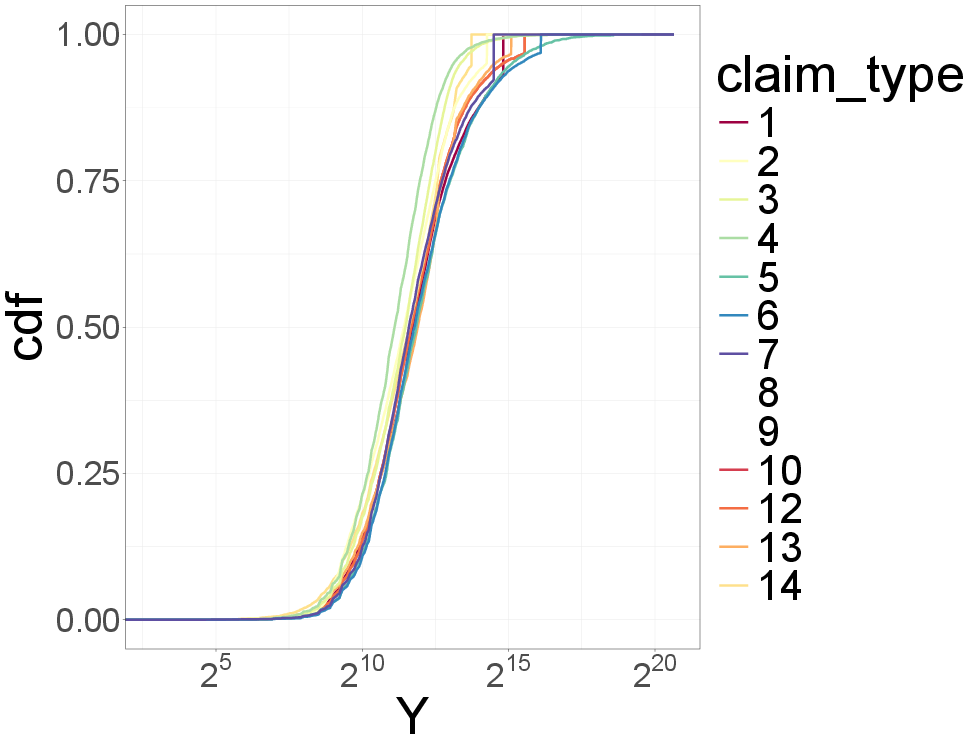}
  \caption{$k=4$.}
  \label{fig:rd1}
\end{subfigure}\hfil
\begin{subfigure}[t]{0.4\linewidth}
  \includegraphics[width=\linewidth]{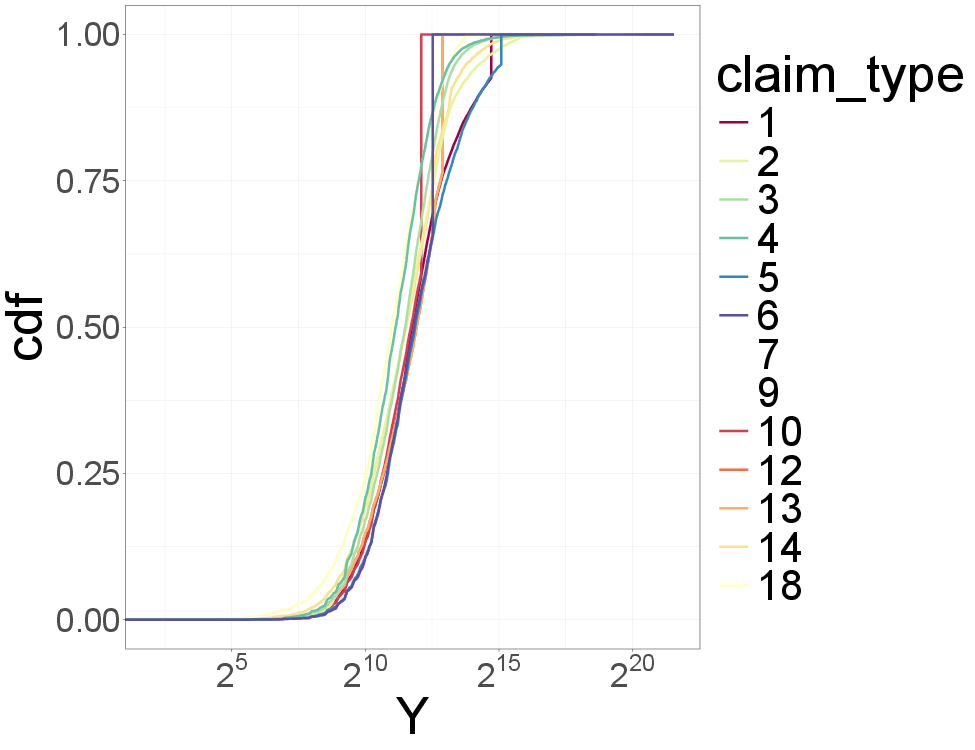}
  \caption{$k=5$.}
  \label{fig:rd2}
\end{subfigure}\hfil
\begin{subfigure}[t]{0.4\linewidth}
  \includegraphics[width=\linewidth]{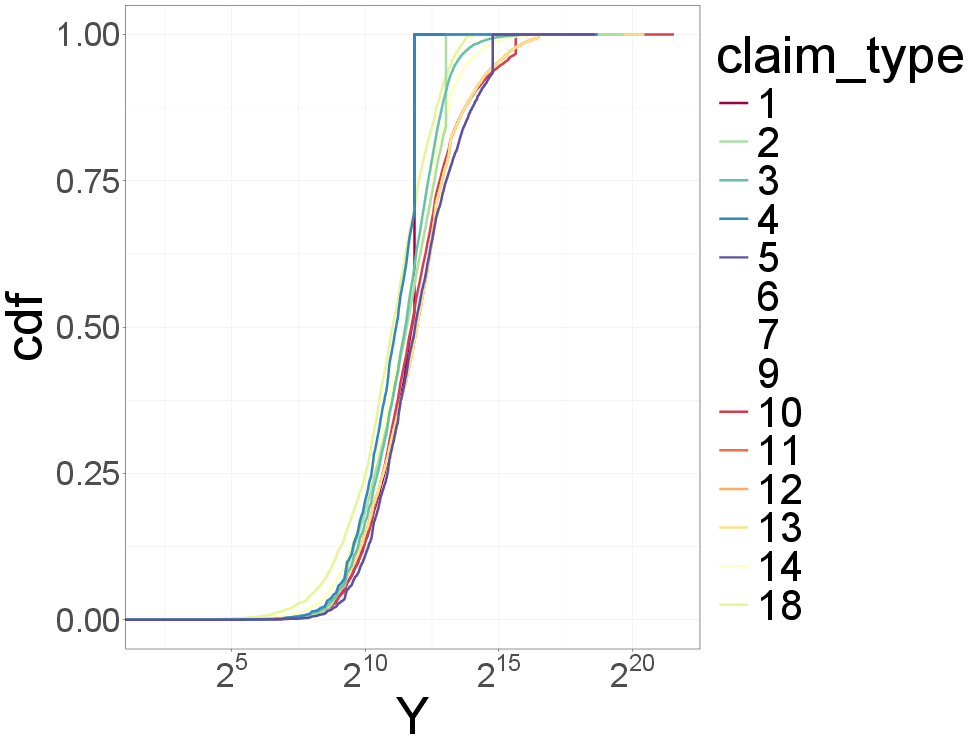}
  \caption{$k=6$.}
  \label{fig:rd3}
\end{subfigure}\hfil
\begin{subfigure}[t]{0.4\linewidth}
  \includegraphics[width=\linewidth]{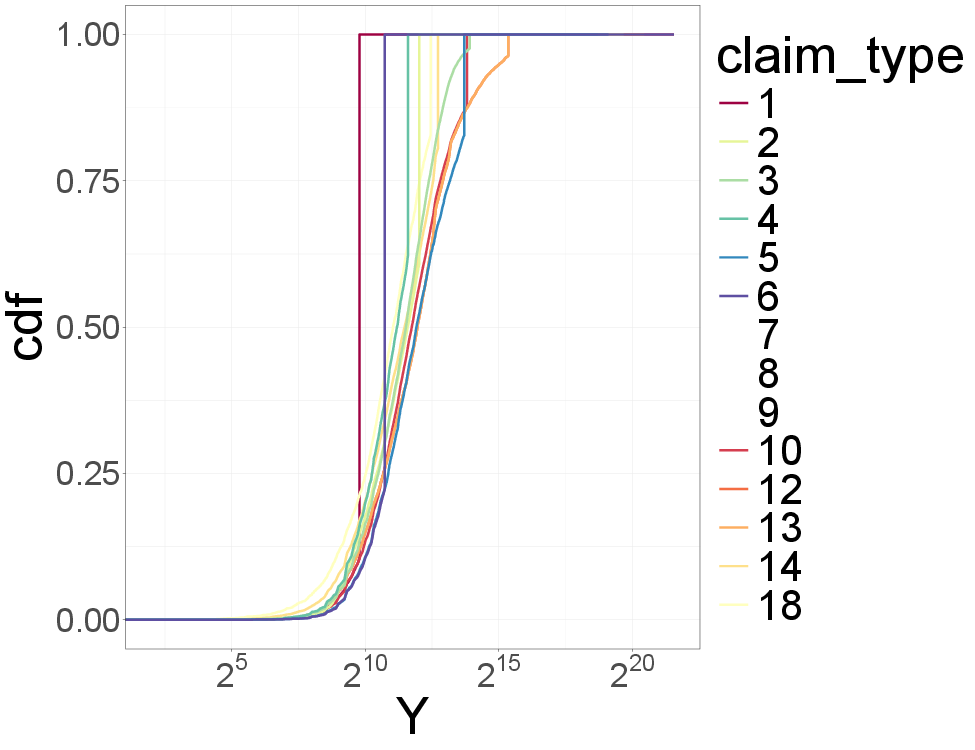}
  \caption{$k=7$.}
  \label{fig:rd4}
\end{subfigure}

\caption{\label{fig:sc_rd} For each different dimension $k=4,5,6,7,8$, we provide the individual claim size curves for our observations by covariate value \texttt{claim\_type}. The x-axis represents $Z$ and we scale it by \texttt{log2} to ease the plot visualization. }
\end{figure}

Consider the first application, where we censor our data at different depths and fit an AJ model using the largest possible state space. Using our model, it is possible to explore and compare the individual claim size curves for the different values of \texttt{claim\_type}, as shown in \Cref{fig:sc_rd}.

The results are reported in \Cref{tab:differentk} and they show that in general the AJ models outperform the CL in terms of \texttt{EI}, except for the scenario with $k=7$. For each scenario, we are able to select a model with or without features using the CRPS measure. The rows with a \cmark (column two) refer to models that include the \texttt{claim\_type} feature in the fit. The target $Y^{\texttt{TOT}}$ is shown in column three. While in other scenarios the CRPS for the AJ model that includes the \texttt{claim\_type} feature is close to or better than the CRPS of the model without the feature, in $k=5$ the model without the feature has a much lower CRPS. The relative variability results show that AJ and CL have comparable results. In fact, only for $k=4$ does CL have a lower relative variability.

\begin{table}[ht]
\centering
\begin{adjustbox}{max width=\textwidth}
\begin{tabular}{rlrrrrrr}
  \hline
$k$ & \texttt{claim\_type} & $Y^{\texttt{TOT}}$  & \texttt{EI} (AJ) & \texttt{EI} (CL) & $\hat{\mbox{sd}}(Y^{\texttt{TOT}})/\hat Y^{\texttt{TOT}}$ (AJ) & $\hat{\mbox{sd}}(Y^{\texttt{TOT}})/\hat Y^{\texttt{TOT}}$(CL) & CRPS (average, relative) \\ 
  \hline
\multirow{2}{.5cm}{4} & \cmark & \multirow{2}{2cm}{616.1327} & -0.0035 &\multirow{2}{1cm}{ 0.0157} & 0.0029 & \multirow{2}{1cm}{0.0023} & 1.0000 \\ 
 & \xmark &  & -0.0029 &  & 0.0029 &  & 1.1403 \\ 
 
\multirow{2}{.5cm}{5} & \cmark &\multirow{2}{2cm}{822.5956 }& -0.0064 & \multirow{2}{1cm}{0.0209} & 0.0008 & \multirow{2}{1cm}{0.0024} & 1.0000 \\ 
 &\xmark &  & -0.0061 &  & 0.0007 &  & 0.4596 \\ 
 
\multirow{2}{.5cm}{6} & \cmark & \multirow{2}{2cm}{999.6005} & -0.0059 & \multirow{2}{1cm}{0.0173} & 0.0017 & \multirow{2}{1cm}{0.0017} & 1.0000 \\ 
 & \xmark &  & -0.0052 & & 0.0017 &  & 0.9987 \\ 
 
\multirow{2}{.5cm}{7}& \cmark & \multirow{2}{2cm}{1190.9112} & -0.0146 & \multirow{2}{1cm}{0.0144} & 0.0011 & \multirow{2}{1cm}{ 0.0017} & 1.0000\\
& \xmark & & -0.0142 &  & 0.0011 &  & 1.0022 \\ 
   \hline
\end{tabular}
\end{adjustbox}
\caption{\label{tab:differentk} For different choices of $k$ (column one), we fit a model with and without \texttt{claim\_type} (column two). The target $Y^{\texttt{TOT}}$ is shown in column three. The \texttt{EI} is shown in columns four and five. The CV of $Y^{\texttt{TOT}}$ are displayed in columns seven and eight for the AJ and CL respectively. The CRPS is shown in column nine.}
\end{table}

For each choice of $k$ we can simply select the best performing model in terms of CRPS and compute the claims reserve, the results are displayed in \Cref{tab:reservestrategyI}. 

\begin{table}[ht]
\centering
\begin{tabular}{rrrl}
  \hline
$k$ & $R$ &  $\mbox{\mbox{sd}}(Y^{\texttt{TOT}})$ & \texttt{claim\_type} \\ 
  \hline
    4 & 10.7832 & 1.8054 & \cmark \\ 
      5 & 7.4981 & 0.5720 & \xmark \\ 
      6 & 10.3268 & 1.6731 & \xmark \\ 
      7 & 5.6317 & 1.3381 & \cmark \\ 
   \hline
\end{tabular}
\caption{\label{tab:reservestrategyI} We selected for each data $k$ (column one), the best performing model in terms of CRPS and present the reserve (column one) and the standard deviation of the reserve (column two). }
\end{table}

\subsection{Model comparison on a single dataset}
\label{ss:modelselection}

In the second application, we censor our data after $5$ accident period and construct the AJ model for $k=4,5,6$. For each choice of $k$, we fit a model both with and without the feature \texttt{claim\_type}. \Cref{tab:strategy1} shows that in terms of CRPS, the model with $k=4$ using the \texttt{claim\_type} feature is the best model (minimum CRPS). This model is also the best in terms of \texttt{EI}.

\begin{table}[ht]
\centering
\begin{adjustbox}{max width=\textwidth}
\begin{tabular}{rlrrrrrr}
  \hline
$k$ & \texttt{claim\_type} & $Y^{\texttt{TOT}}$& \texttt{EI} (AJ) & \texttt{EI} (CL) & $\hat{\mbox{sd}}(Y^{\texttt{TOT}})/\hat Y^{\texttt{TOT}}$ (AJ) & $\hat{\mbox{sd}}(Y^{\texttt{TOT}})/\hat Y^{\texttt{TOT}}$(CL) & CRPS (average, relative)\\ 
  \hline
\multirow{2}{.5cm}{4} &\xmark&\multirow{6}{2cm}{999.6005} & -0.0059 & \multirow{6}{1cm}{0.0173} & 0.0016 & \multirow{6}{1cm}{0.0017} & 1.0143 \\
  &\cmark&& -0.0040 &  & 0.0020 &  & 1.0000 \\
\multirow{2}{.5cm}{5} &\xmark&& -0.0069 &  & 0.0015 &  &  0.9916 \\  
&\cmark&& -0.0052 &  & 0.0018 &  & 1.0000 \\  
\multirow{2}{.5cm}{6} &\xmark&& -0.0052 &  & 0.0017 &  &0.9987 \\ 
&\cmark&& -0.0059 &  & 0.0017 &  & 1.0000\\ 
   \hline
\end{tabular}
\end{adjustbox}
\caption{\label{tab:strategy1} For the dataset with depth $5$ accident periods, we fit the AJ model both including and excluding the \texttt{claim\_type} feature (column two). The target $Y^{\texttt{TOT}}$ is displayed in column three. The \texttt{EI} for the AJ and the CL can be found in columns four and five. The coefficient of variation of $Y^{\texttt{TOT}}$ is displayed in columns six and seven. The CRPS is displayed in column eight.}
\end{table}

Using the CRPS we are able to select the model with $k=4$ and that uses \texttt{claim\_type} as a feature to compute the claims reserve. The model provides a reserve of $11.485$ millions and a standard deviation of $ 2.0385$ millions. The strategy presented in this last section is of particular interest to reserving actuaries. While the empirical study in \Cref{ss:differentk} is interesting for understanding the behaviour of our model on different data, in practice the maximum depth of the data is selected using indicators such as claims settlement speed or using the expert judgement of an experienced actuary \citet[p.~12]{wuthrich08}. The depth of the data is then known and the main interest is simply to select the best model, as shown here. Selecting the most appropriate model can be very difficult, see again \citet[p.~13]{wuthrich08}, and the strategy presented in this section provides a mathematically sound approach to doing so.

\section{Conclusion}\label{sec:conc}

The methodology described in this manuscript presents an improved approach for predicting loss reserves compared to commonly used aggregate loss reserving methods. We also introduce a comprehensive non-parametric estimator for the cumulative density function of individual claim sizes, distinguishing it from other individual loss reserving methods that mainly focus on point forecasts of claim reserves.
While some questions remain open, particularly regarding the modeling of Incurred But Not Reported (IBNR) claims and their dependence on specific features, practical applications can address these issues by fitting our model with the chain ladder method for future reports and projecting their future costs using the unconditional curve of individual claim sizes.
Furthermore, this methodology can be extended in various ways. One possibility is to explore the use of reverse time models, as described in \citet{hiabu17}, to project the future exposure of IBNR claims. In a broader context, future research could investigate how to model not only the size but also the frequency of individual claims.

\section*{Supplementary material}

The code at \href{https://github.com/gpitt71/conditional-aj-reserving.git}{gpitt71/conditional-aj-reserving} complements the results of this manuscript. The folder was registered with unique Zenodo DOI \href{https://zenodo.org/doi/10.5281/zenodo.10118895}{10.5281/zenodo.10118895}.

\section*{Competing interests}
The authors declare no competing interests of any kind.

\printbibliography

\appendix

\section{Comparison with \texttt{hirem}}
\label{appendix:hirem}

We perform a first comparison with the \texttt{hirem} model on the scenario Beta that we introduced in \Cref{ss:empirical_analysis}. The results are reported in \Cref{tab:hirem_beta}. For the different values of $k$ (column one), we compare the AJ model (with feature $U$) to the CL and \texttt{hirem} in terms of \texttt{EI} and CRPS. Each row corresponds to a different model (column two). The (average) target $Y^{\texttt{TOT}}$ is reported in column three. The \texttt{EI} (column four) shows that the AJ model is predicting the target $Y^{\texttt{TOT}}$ more accurately than the CL and \texttt{hirem}. We also observe that the \texttt{hirem} model seems to be less accurate for higher values of $k$. Consistently with the data generating assumptions, the CRPS indicates that the AJ model better describes the curve of the claim size. The relative variation is reported from \Cref{tab:sim_beta} for the AJ model and the CL model (column five). We observe a similar relative variation of $Y^{\texttt{TOT}}$ on our models compared to \texttt{hirem}.

\begin{table}[ht]
\centering
\begin{tabular}{rlrrrr}
  \hline
Scenario & Model & $Y^{\texttt{TOT}}$ (average) & \texttt{EI} & $\hat{\mbox{sd}}(Y^{\texttt{TOT}})/\hat Y^{\texttt{TOT}}$ & CRPS (average, relative) \\ 
  \hline
\multirow{3}{1.1cm}{Beta (k=4)} & AJ & \multirow{3}{1.1cm}{42.279} & 0.006 & 0.008 & 1.000 \\ 
   & CL &  & 0.103 & 0.008 & - \\ 
   & \texttt{hirem} &  & -0.028 & 0.009 & 1.582 \\
   \hline
  \multirow{3}{1.1cm}{Beta (k=5)} & AJ & \multirow{3}{1.1cm}{94.580 }& 0.011 & 0.009 & 1.000 \\
   & CL &  & 0.190 & 0.008 & - \\ 
   & \texttt{hirem} &  & 0.071 & 0.009 & 1.680 \\
   \hline
  \multirow{3}{1.1cm}{Beta (k=6)} & AJ & \multirow{3}{1.1cm}{108.423} & -0.013 & 0.007 & 1.000 \\ 
   & CL &  & 0.256 & 0.008 & - \\ 
   & \texttt{hirem} &  & 0.148 & 0.009 & 1.697 \\ 
   \hline
  \multirow{3}{1.1cm}{Beta (k=7)} & AJ & \multirow{3}{1.1cm}{116.032} & -0.033 & 0.007 & 1.000 \\  
   & CL &  & 0.255 & 0.008 & - \\ 
   & \texttt{hirem} &  & 0.168 & 0.008 & 1.790 \\ 
   \hline
\end{tabular}
\caption{ \label{tab:hirem_beta} Results for scenario Beta. For each value of $k$ (column one) we present the average results over the $20$ simulations. Each row of the table corresponds to a different model (column two). The table includes the (average) actual simulated total cost (column three) and the error incidence for the AJ, the CL, and \texttt{hirem} (columns four). In columns five we show the average relative variation and in column six we reported the average CRPS relative to the AJ CRPS.}

\end{table}

A second comparison is performed on synthetic data generated from the individual claims simulators available in the R package \citet{hirempackage}. The package includes the \texttt{hirem} models implementation and four individual claims simulators for reported claims for $k=10$. While the four scenarios are briefly introduced in this section, we refer to the package documentation and the main manuscript for a detailed description. The generated features that are relevant for our application are reported in \Cref{tab:hirem}.

The data include one categorical feature, the type of claim (\texttt{Type}), which we use in our application and a hidden feature (\texttt{Hidden}) that is not known to the insurer at the evaluation. The package includes a baseline simulated scenario (Baseline) and three scenarios that include modifications compare to the Baseline. In the scenario Claim Mix, there is a change in the portfolio distribution with respect to the feature \texttt{Type}. In the scenario Extreme, a seasonality effect in calendar year $9$ simulates an increase in the claim occurrences. Lastly, the Settlement scenario simulates a change of the claim settlement process to quicker settlement after calendar period $7$.

The results of this second application are reported \Cref{tab:hirem_hirem}. We find that, both in terms of \texttt{EI} and CRPS and in the four simulated scenarios, our models always outperform the CL, while \texttt{hirem} performs better than the AJ (columns five and seven). The CRPS in column seven are relative to the CRPS of the AJ model with features. These results are expected, as the data are generated using assumptions that are consistent with the \texttt{hirem} model, see Appendix A of \citet{crevecoeur23}. Notwithstanding, the AJ model is performing comparatively well. The results are averaged over ten simulations.

We remark that the applications reported in \Cref{tab:hirem_beta} and \Cref{tab:hirem_hirem} indicate that, in general, for the scenarios that we inspected, individual models always outperform the chain ladder benchmark. However, as expected, the models perform best when the data generating process is consistent with the models assumptions.

\begin{table}[ht]
\centering
\begin{tabular}{l|l}
 Covariates & Description\\ \hline
\texttt{claim.nr} & Policy identifier\\
 \texttt{Type} $\in  \left\{T1, T2, T3 \right\}$ & Type of claim \\
 \texttt{Hidden} $\in  \left\{I, M, H \right\}$ & Covariate unknown to the insurer (disregarded for modeling) \\
\texttt{occ.year} & Accident year\\
\texttt{rep.year} & Calendar year of reporting\\
\texttt{dev.year} & Development year\\
\texttt{size} & Incremental paid amount \\
\texttt{settlement} & Indicator, $1$ in the development year of settlement \\
 \hline
  \end{tabular}
  \caption{\label{tab:hirem} Description of the \texttt{hirem} data.  }
\end{table}



\begin{table}[ht]
\centering

\begin{tabular}{lllrrrr}
  \hline
Scenario & Type & Model & $Y^{\texttt{TOT}}$(average) & \texttt{EI} & $\hat{\mbox{sd}}(Y^{\texttt{TOT}})/\hat Y^{\texttt{TOT}}$ & CRPS (average, relative) \\ 
  \hline
\multirow{4}{1.1cm}{Baseline}  & \cmark & \multirow{2}{1.1cm}{AJ}  & \multirow{4}{1.1cm}{50.240} & 0.070 & 0.004 &  1.000 \\
   & \xmark &  &  & 0.067 & 0.004 & 1.001 \\
   & \cmark & \texttt{hirem} & & 0.012 & 0.002 &  0.785 \\  
   & \xmark & CL &  & 0.232 & 0.010 &   \\ 
   \hline
  \multirow{2}{1.1cm}{Claim Mix} & \cmark & \multirow{2}{1.1cm}{AJ} & \multirow{4}{1.1cm}{57.880}  & 0.071 & 0.004 &  1.000 \\  
   & \xmark &  &  & 0.066 & 0.004&  1.002 \\
   & \cmark & \texttt{hirem} &  & 0.014 & 0.002 &  0.792 \\  
   & \xmark & CL &  & 0.240 & 0.010 &  \\ 
   \hline
  \multirow{2}{1.1cm}{Extreme} & \cmark & \multirow{2}{1.1cm}{AJ} & \multirow{4}{1.1cm}{48.812 }   & 0.076 & 0.005 &  1.000 \\  
   & \xmark &  &  & 0.069 & 0.004 &  1.002 \\ 
   & \cmark & \texttt{hirem} &   & 0.039 & 0.002 &  0.755 \\ 
   & \xmark & CL &  & 0.231 & 0.010 &  \\ 
  \hline
  \multirow{2}{1.1cm}{Settlement} & \cmark & \multirow{2}{1.1cm}{AJ} & \multirow{4}{1.1cm}{50.241} & 0.069 & 0.004 &  1.000 \\ 
   & \xmark &  &  & 0.065 & 0.004 & 1.001 \\
   & \cmark & \texttt{hirem} &  &0.011 & 0.002 &  0.766 \\
   & \xmark & CL &  & 0.231 & 0.001 &  \\
   \hline
\end{tabular}
\caption{\label{tab:hirem_hirem}The \texttt{hirem} package includes four data generators in four different scenarios with $k=10$ (column one). We compare our models (AJ with and without features, column two) to the model in \citet{crevecoeur23} and the CL (column three). The actual $Y^{\texttt{TOT}}$ target is reported in column four. We show the \texttt{EI} and the CRPS results in columns five and seven respectevely. The predicted relative variation of $Y^{\texttt{TOT}}$ is shown in column six.}
\end{table}

\end{document}

%% file: preamble.tex
\usepackage{geometry,setspace}
\small\normalsize
\usepackage{hyperref}
\hypersetup{
    colorlinks=true,
    citecolor=teal,
    linkcolor=teal
}
\usepackage{caption}
\usepackage{enumerate}
\usepackage{graphicx}
\usepackage{rotating,multirow}
\usepackage[table]{xcolor}
 \usepackage{amsmath,amsfonts,amssymb,mathrsfs,amsthm, yhmath}

\DeclareMathOperator{\Var}{Var}
 \usepackage{mathtools}
\usepackage{bbm}
\usepackage{bm}
\usepackage[style=ext-authoryear,maxbibnames=9,maxcitenames=2,uniquename=init,autocite=inline,giveninits=true,backend=biber, natbib]{biblatex}
\DeclareNameAlias{author}{family-given}
\DeclareFieldFormat[article,periodical]{volume}{\mkbibbold{#1}}

\usepackage{booktabs}
\usepackage{adjustbox}

\usepackage{pdflscape}
\usepackage{todonotes}
\usepackage{authblk}
\usepackage{xr}
\usepackage{tikz-cd}
\usepackage{framed}

\usepackage{xfrac}

\usepackage{caption}
\usepackage[T1]{fontenc}
\input{glyphtounicode}
\usepackage{tikz}
\usepackage{ subcaption, cleveref}
\usepackage{soul}
\usepackage{prodint}
\usepackage{stackengine}
\newcommand{\indep}{\perp \!\!\! \perp}
\usepackage{pifont}   
\newcommand{\xmark}{\ding{55}}
\newcommand{\cmark}{\ding{51}}

\usetikzlibrary{arrows,positioning,calc} 
\usetikzlibrary{decorations.pathreplacing}
\tikzset{
    >=stealth',
    punkt/.style={
           rectangle,
           rounded corners,
           draw=black, thick,
           text width=5.5em,
           minimum height=2em,
           text centered},
    punktl/.style={
           rectangle,
           rounded corners,
           draw=black, thick,
           text width=7em,
           minimum height=2em,
           text centered},
    pil/.style={
           ->,
           shorten <=4pt,
       shorten >=4pt
    },
    pildotted/.style={
           ->,
           shorten <=4pt,
           shorten >=4pt,
  dotted,
  },
    punktf/.style={
           rectangle,
           text width=4.0em,
           minimum height=1.5em,
           text centered},
    punktfTop/.style={
           rectangle,
           text width=4.0em,
           minimum height=1.5em,
           text centered,
           append after command={
               [thick,shorten >=0.2bp, shorten <=0.2bp]
               (\tikzlastnode.north west)edge(\tikzlastnode.north east)
}
    },
    punktfBot/.style={
           rectangle,
           text width=4.0em,
           minimum height=1.5em,
           text centered,
           append after command={
               [thick,shorten >=0.2bp, shorten <=0.2bp]
               (\tikzlastnode.south west)edge(\tikzlastnode.south east)
            }
    }
}

\theoremstyle{definition}
\newtheorem{definition}{Definition}
 \newtheorem{assumption}{Assumption}
\newtheorem{remark}{Remark}

\usepackage{mathbbol}
\DeclareSymbolFontAlphabet{\amsmathbb}{AMSb}%

\usepackage{accents}

\renewcommand{\leq}{\leqslant}

\renewcommand{\le}{\leqslant}
\renewcommand{\ge}{\geqslant}